\documentclass[journal]{IEEEtran}
\usepackage[cmex10]{amsmath}
\usepackage{amssymb,amsthm}
\usepackage{url}
\usepackage[dvips]{graphicx}
\usepackage{psfrag}
\usepackage{stfloats}
\allowdisplaybreaks[3]

\newtheorem{theorem}{Theorem}
\newtheorem{lemma}{Lemma}

\def\Pb{\mathrm{P_b}}
\def\Pn{\mathbb{P}_n}

\title{Effects of Single-Cycle Structure on Iterative Decoding for Low-Density Parity-Check Codes}

\author{\IEEEauthorblockN{Ryuhei Mori\IEEEauthorrefmark{1}, Toshiyuki Tanaka\IEEEauthorrefmark{1}, Kenta Kasai\IEEEauthorrefmark{2},
and Kohichi Sakaniwa\IEEEauthorrefmark{2}}
\thanks{
\IEEEauthorrefmark{1}R. Mori and T. Tanaka are with the Department of System Science, Graduate School of Informatics, Kyoto University, Kyoto, 606-8501, Japan (e-mail: {rmori@sys.i.kyoto-u.ac.jp}, {tt@i.kyoto-u.ac.jp})}
\thanks{\IEEEauthorrefmark{2}K. Kasai and K. Sakaniwa are with the Department of Communications and Integrated Systems, Graduate School of Science and Engineering,
Tokyo Institute of Technology, Tokyo, 152-8552, Japan
(e-mail: {\{kenta,sakaniwa\}@comm.ss.titech.ac.jp})}
}

\begin{document}
\maketitle

\begin{abstract}
We consider communication over the binary erasure channel (BEC) using low-density parity-check (LDPC) codes and belief propagation (BP) decoding.
For fixed numbers of BP iterations, the bit error probability approaches a limit as blocklength tends to infinity,
and the limit is obtained via density evolution.
On the other hand, the difference between the bit error probability of codes with blocklength $n$ and that in the large blocklength limit is
asymptotically $\alpha(\epsilon,t)/n + \Theta(n^{-2})$
where $\alpha(\epsilon,t)$ denotes a specific constant determined by the code ensemble considered, the number $t$ of iterations, and the erasure probability $\epsilon$ of the BEC.
In this paper, we derive a set of recursive formulas which allows evaluation of the constant $\alpha(\epsilon,t)$ for standard irregular ensembles.
The dominant difference $\alpha(\epsilon,t)/n$ can be considered as effects of cycle-free and single-cycle structures of local graphs.
Furthermore, it is confirmed via numerical simulations that
estimation of the bit error probability using $\alpha(\epsilon,t)$ is accurate even for small blocklengths.
\end{abstract}

\section{Introduction}
\IEEEPARstart{I}{t} is well known that
low-density parity-check (LDPC) codes for transmission over binary memoryless symmetric channels
approach channel capacity with low-complexity iterative decoder called \textit{belief propagation} (BP) decoder.
Especially, for the binary erasure channels (BEC), LDPC codes with BP decoder provably achieve channel capacity~\cite{258573}.
Large-blocklength limit of the bit error probability 
of BP decoder with a fixed number of iterations 
can be calculated by the method called
\textit{density evolution}~\cite{910577}.
In this paper, we consider how fast the bit error probability approaches the limit as blocklength tends to infinity.
Although performance analysis of LDPC codes is often developed for 
general binary-input memoryless symmetric channels~\cite{910577,Ri03,Mon06,DBLP:journals/corr/abs-cs-0511039,4557356}, 
we restrict our attention in this paper to the case where the channel is the BEC, 
since performance analysis on the BEC~\cite{258573,1003839,1715529,4675720,4777618} 
is generally simpler than that for general channels. 
In density evolution,
the bit error probability is calculated recursively
by considering tree neighborhoods whose depth is equal to the number of iterations.
In the analysis of this paper, we consider not only tree neighborhood graphs but also single-cycle neighborhood graphs
in order to derive the most dominant term in the bit error probability which vanishes in the large-blocklength limit.  
We would like to mention that it might be possible to generalize our analysis to other channels and iterative decoders
since the approach taken in our analysis is based on density evolution which is applicable to any combination of a channel and an iterative decoder.

Let $\Pb(n,\epsilon,t)$ denote the bit error probability of an ensemble of codes of blocklength $n$ over the BEC($\epsilon$) after $t$ BP iterations.
The large-blocklength limit of the bit error probability after $t$ iterations is denoted by $\Pb(\infty,\epsilon,t)$.
Evaluation of $\Pb(\infty,\epsilon,t)$ using density evolution
has revealed that there exists a threshold erasure probability $\epsilon_\mathrm{BP}$ such that
the bit error probability $\Pb(\infty,\epsilon,t)$ after a sufficient number of BP iterations tends to $0$ if $\epsilon<\epsilon_\mathrm{BP}$ 
and to a strictly positive value if $\epsilon>\epsilon_\mathrm{BP}$. 

From a practical point of view, it is desirable to evaluate $\Pb(n, \epsilon, t)$ 
for a finite $n$, which, however, is much more complicated than the evaluation of $\Pb(\infty, \epsilon, t)$. 
The bit and block error probabilities for finite blocklength and for infinite number of iterations 
are calculated exactly via stopping-set analysis for regular ensembles~\cite{1003839}
and also for irregular ensembles~\cite{1023275}.
Furthermore, the bit and block error probabilities of expurgated ensembles for finite blocklength and for finite numbers of iterations are also calculated exactly in a combinatorial way~\cite{RiU02/LTHC}.
However, these analyses require high computational costs which grow like a power of the blocklength and like an exponential of the number of degrees.
This fact severely restricts usefulness of these analyses. 

An approach to a finite-length analysis for irregular ensembles with low computational complexity would be to consider large-$n$ asymptotics. 
There are two efficient methods to derive large-$n$ asymptotics
for the bit error probability for blocklength $n$ and for infinite number of iterations, 
which is denoted by $\Pb(n,\epsilon,\infty)$.  
The method proposed by Di, Richardson, and Urbanke~\cite{1715529} has shown that 
the bit error probability below the threshold after infinite number of iterations is expressed as
\begin{equation}
\Pb(n,\epsilon,\infty) = \frac1{2}\frac{\epsilon\lambda'(0)\rho'(1)}{1-\epsilon\lambda'(0)\rho'(1)}\frac1{n} + o\left(\frac1{n}\right). \label{eq:appinf}
\end{equation}
One may thus obtain an approximation formula for $\Pb(n,\epsilon,\infty)$ by ignoring 
the term $o(n^{-1})$ in \eqref{eq:appinf}.  
However, the approximation is not accurate near the threshold
for any irregular ensembles due to the following reasons.
If the limit $\lim_{n\to\infty}\Pb(n,\epsilon,\infty)$ is discontinuous at $\epsilon_\mathrm{BP}$ as a function of $\epsilon$
(i.e., $\lim_{n\to\infty}\Pb(n,\epsilon_\mathrm{BP},\infty)>0$), convergence to the limit is not uniform 
since $\Pb(n,\epsilon,\infty)$ for any finite $n$ is continuous with respect to $\epsilon$.  
Hence, an arbitrarily large blocklength is required near the threshold so that the above approximation formula is expected to be accurate.  
On the other hand, 
the convergence is uniform for $\epsilon\in[0,\epsilon_\mathrm{BP}]$ 
if the limit $\lim_{n\to\infty}\Pb(n,\epsilon,\infty)$ is continuous at $\epsilon_\mathrm{BP}$ as a function of $\epsilon$
(i.e., $\lim_{n\to\infty}\Pb(n,\epsilon_\mathrm{BP},\infty)=0$).  
In such cases, however, 
the coefficient of $n^{-1}$ in~\eqref{eq:appinf} diverges as $\epsilon$ approaches the threshold $\epsilon_\mathrm{BP}$ from below, 
since the threshold is given as $\epsilon_\mathrm{BP}=(\lambda'(0)\rho'(1))^{-1}$.  
Hence, an arbitrarily large blocklength is again required near the threshold 
so that the above-mentioned approximation formula is expected to be accurate.  
From the above facts, the approximation \eqref{eq:appinf} is accurate only for a small-$\epsilon$ region which is often called an error floor.

As an alternative approach, a method that is based on scaling law has been proposed~\cite{4777618,Am07}, 
which requires only a constant cost and is useful for estimation of the bit and block error probabilities near the threshold
where the error probabilities behave like what is called a waterfall curve. 
This analysis permits finite-length optimization which maximizes rate of a code under a given blocklength, erasure probability and allowable error probability.

Both of these two methods are, however, 
applicable only for infinite number of iterations, 
whereas the number of iterations is often constrained in practical applications due to limitation of resources, 
e.g., time, energy, etc., 
so that results for finite numbers of iterations should be more significant than those for infinite number of iterations.  
We therefore focus in this paper on an asymptotic bit error probability with respect to the blocklength when the number $t$ of iterations is finite and fixed.
The basic idea underlying our approach is to consider a large-$n$ asymptotic expansion of the bit error probability 
and to evaluate the second dominant term in the asymptotic expansion. 
There exists a coefficient $\alpha(\epsilon,t)$ of $n^{-1}$ on the basis of which the asymptotic expansion of $\Pb(n,\,\epsilon,\,t)$ is expressed as
\begin{equation}
\Pb(n,\epsilon,t) = \Pb(\infty,\epsilon,t) + \alpha(\epsilon,t)\frac1{n} + o\left(\frac1{n}\right)\text{.}\label{eq:apx}
\end{equation}
The second term $\alpha(\epsilon,t)/n$ in the right-hand side of~(\ref{eq:apx}) 
is determined by tree and single-cycle structures of local graphs,
while the first term $\Pb(\infty,\epsilon,t)$ is due to only tree local graphs.
An important consequence of considering a finite-$t$ asymptotic expansion is that 
the approximation formula derived by ignoring the term $o\left(n^{-1}\right)$ in~(\ref{eq:apx}) is expected to be accurate for all $\epsilon$ uniformly
if the blocklength is sufficiently large, 
since the convergence $\lim_{n\to\infty}\Pb(n,\epsilon,t)$ is uniform for $\epsilon\in[0,1]$,
as we will see in later sections.  
Our main result is to derive a set of recursive formulas which allows evaluation of the coefficient $\alpha(\epsilon,t)$ for irregular ensembles.



In Section~\ref{sec:prel}, we define random ensembles of graphs used in this paper.
In Subsection~\ref{subsec:dec}, we see how the coefficient $\alpha(\epsilon,t)$ is decomposed
into two components, one representing contributions of cycle-free neighborhood graphs 
and the other representing contributions of single-cycle neighborhood graphs.
In Subsection~\ref{subsec:beta}, we obtain the component for cycle-free neighborhood graphs in $\alpha(\epsilon,t)$ 
by developing a generating function method.
In Subsection~\ref{subsec:coef}, we see how to enumerate the coefficient of $n^{-1}$ in 
asymptotic expansion of the probability for single-cycle neighborhood graphs.  
The technique developed in Subsection~\ref{subsec:coef} is then used in the calculation of the contribution of single-cycle neighborhood graphs
in Subsection~\ref{subsec:gamma} via the single-cycle neighborhood graph ensemble defined in Subsection~\ref{subsec:scnge}.
%
In Section~\ref{sec:limalpha}, we study the limit $\lim_{t\to\infty}\alpha(\epsilon,t)$ for regular ensembles.
In Section~\ref{sec:uniconv}, we show that when the number of iterations is fixed, the large blocklength convergence is uniform with respect to $\epsilon$.
It implies that, for sufficiently large blocklength, the approximation~\eqref{eq:apx} is uniformly accurate for all $\epsilon$.
Furthermore, 
in Section~\ref{sec:cal}, it is confirmed via numerical simulations that the approximations for several ensembles are accurate even for small blocklength.
Finally, we conclude this paper in Section~\ref{sec:conc}.
The results of this paper have also been presented in conference papers~\cite{4595026,4658691,5206063}.

\section{Preliminaries}\label{sec:prel}
\subsection{Tanner graphs}
A Tanner graph $G=(\mathcal{V},\mathcal{C},\mathcal{S_V},\mathcal{S_C},\mathcal{E})$
is a bipartite graph which is represented by a set $\mathcal{V}$ of variable nodes, a set $\mathcal{C}$ of check nodes,
a set $\mathcal{S_V}$ of variable-node sockets, a set $\mathcal{S_C}$ of check-node sockets, 
and a set $\mathcal{E}$ of edges connecting variable-node and check-node sockets. 
A node $m$ is identified as a tuple of sockets associated with $m$.
To be precise,
\begin{align*}
\mathcal{V} &\subseteq \bigcup_{k=1}^\infty \{(s_1,\dotsc,s_k)\mid s_1\in \mathcal{S_V}, \dotsc, s_k\in \mathcal{S_V}\}\\
\mathcal{C} &\subseteq \bigcup_{k=1}^\infty \{(t_1,\dotsc,t_k)\mid t_1\in \mathcal{S_C}, \dotsc, t_k\in \mathcal{S_C}\}\text{.}
\end{align*}
Sockets associated with the same node are all distinct.
Each variable-node socket $s\in \mathcal{S_V}$ is associated with one and only one node in $\mathcal{V}$.  
Similarly, each check-node socket $t\in \mathcal{S_C}$ is associated with one and only one node in $\mathcal{C}$. 
An edge $e$ is identified as a pair of sockets which connect to $e$.
To be precise,
\begin{align*}
\mathcal{E} \subseteq \{(s,t)\mid s\in \mathcal{S_V}, t\in \mathcal{S_C}\}\text{.}
\end{align*}
Each socket connects to one and only one edge.
The number of variable-node sockets, the number of check-node sockets and the number of edges are equal, 
i.e.,\ $|\mathcal{S_V}|=|\mathcal{S_C}|=|\mathcal{E}|$.

\subsection{Irregular LDPC code ensembles}
\label{subsec:irregens}
An $(n,L(x),R(x))$-\textit{irregular ensemble} is a random ensemble of LDPC codes of blocklength $n$ which are represented by Tanner graphs
with variable-node degree distribution polynomial $L(x)$ from node perspective and check-node degree distribution polynomial $R(x)$ from node perspective~\cite{258573}.
These two degree distribution polynomials are expressed as
\begin{align*}
L(x) &:= \sum_i L_ix^{i},& R(x) &:= \sum_i R_jx^{j}\text{.}
\end{align*}
Each Tanner graph in the $(n,L(x),R(x))$-irregular ensemble has $n$ variable nodes, fraction $L_i$ of variable nodes of degree $i$, 
and fraction $R_j$ of check nodes of degree $j$.
The sets $\mathcal{V}$, $\mathcal{C}$, $\mathcal{S_V}$ and $\mathcal{S_C}$ 
defining nodes and sockets in the Tanner graphs are arbitrarily fixed in an ensemble.
Each instance of the edge set $\mathcal{E}$ is chosen randomly from all $E!$ possible realizations 
with uniform probability, where $E:=|\mathcal{E}|=nL'(1)$ is the number of edges of the Tanner graphs.  

We also use degree distribution polynomials $\lambda(x)$ and $\rho(x)$ from edge perspective, which are defined as
\begin{align*}
\lambda(x) &= \sum_i\lambda_ix^{i-1} := \frac{L'(x)}{L'(1)}\\
\rho(x) &= \sum_j\rho_jx^{j-1} := \frac{R'(x)}{R'(1)}\text{.}
\end{align*}
An $(n,L(x),R(x))$-irregular ensemble has fraction $\lambda_i$ of edges incident to a variable node of degree $i$ and
fraction $\rho_j$ of edges incident to a check node of degree $j$.
An $(n,L(x),R(x))$-irregular ensemble is also referred to as an $(n,\lambda(x),\rho(x))$-irregular ensemble.

\textit{The bit error probability of} $(n,\lambda(x),\rho(x))$-\textit{irregular ensemble}
is defined as the average bit error probability of 
instance codes.
In this paper, we deal with an asymptotic bit error probability with respect to blocklength
with the fixed degree distributions $(\lambda(x),\,\rho(x))$.
In the following, we will also use the notation $(\lambda(x),\,\rho(x))$ 
to specify an irregular ensemble when the blocklength is not relevant. 


\subsection{Neighborhood graph ensembles}
\textit{Distance} of two nodes in a Tanner graph is defined as the number of check nodes in the shortest path between the two nodes except both ends.
A \textit{neighborhood graph} of depth $t$ of a variable node $v_0$ is
a subgraph which consists of the variable nodes and the check nodes with distance from $v_0$ not greater than $t$ and $(t-1)$, respectively.
Each neighborhood graph is expressed as $G=(\mathcal{V_N}, \mathcal{C_N}, v_0, \mathcal{S_{V_N}}, \mathcal{S_{C_N}}, \mathcal{E_N})$.
Sets $\mathcal{V_N}$, $\mathcal{C_N}$, $\mathcal{S_{V_N}}$, $\mathcal{S_{C_N}}$ and $\mathcal{E_N}$
are a set of variable nodes, a set of check nodes, a set of variable-node sockets, 
a set of check-node sockets and a set of edges, respectively.
Expressions and roles of $\mathcal{V_N}$, $\mathcal{C_N}$, $\mathcal{S_{V_N}}$, $\mathcal{S_{C_N}}$ and $\mathcal{E_N}$
are the same as 
those of $\mathcal{V}$, $\mathcal{C}$, $\mathcal{S_{V}}$, $\mathcal{S_{C}}$ and $\mathcal{E}$ for a Tanner graph, respectively.
A variable node $v_0\in \mathcal{V_N}$ is called the \textit{root node}.
\textit{Depth} of a node in $G$ is the distance from the root node $v_0$.
Variable nodes of depth $t$ may have sockets which do not connect to any edges.
With an abuse of notations, we will also write $\mathcal{V}(G) := \mathcal{V_N}$ and $\mathcal{C}(G) := \mathcal{C_N}$.

A \textit{neighborhood graph ensemble} $\mathcal{N}_t(n,\lambda(x),\rho(x))$ 
induced by an $(n,\lambda(x),\rho(x))$-irregular ensemble 
is an ensemble of neighborhood graphs of depth $t$.
Each neighborhood graph $G$ 
is associated with the probability $\mathbb{P}_n(G)$ which is defined by the following steps.
We first consider a Tanner graph $(\mathcal{V},\mathcal{C},\mathcal{S_V},\mathcal{S_C},\mathcal{E})$ generated from 
the $(n,\lambda(x),\rho(x))$-irregular ensemble. 
For a neighborhood graph $G=(\mathcal{V_N}, \mathcal{C_N}, v_0, \mathcal{S_{V_N}}, \mathcal{S_{C_N}}, \mathcal{E_N})$ 
with $\mathcal{V_N}\subseteq\mathcal{V}$, $\mathcal{C_N}\subseteq\mathcal{C}$, $\mathcal{S_{V_N}}\subseteq\mathcal{S_V}$,
$\mathcal{S_{C_N}}\subseteq\mathcal{S_C}$ and $\mathcal{E_N}\subseteq\mathcal{E}$, 
where the root node $v_0$ is chosen uniformly from $\mathcal{V}$, 
and where a variable node $v\in \mathcal{V}$ and a check node $c\in \mathcal{C}$ are members of $\mathcal{V_N}$ and $\mathcal{C_N}$
if and only if their distances from $v_0$ are not greater than $t$ and $(t-1)$, respectively.
Similarly, a variable-node socket $s\in \mathcal{S_V}$ and a check-node socket $t\in \mathcal{S_C}$ are members of $\mathcal{S_{V_N}}$ and $\mathcal{S_{C_N}}$
if and only if $s$ and $t$ are associated with nodes in $\mathcal{V_N}$ and $\mathcal{C_N}$, respectively.
An edge $(s,t)\in\mathcal{E}$ is a member of $\mathcal{E_N}$ if and only if $s\in \mathcal{S_{V_N}}$ and $t\in \mathcal{S_{C_N}}$.

The random choice of the edge set $\mathcal{E}$ in the original irregular ensemble 
induces a probability distribution over the set of neighborhood graphs, 
under which each possible neighborhood graph $G$ has a probability
\begin{equation*}
\mathbb{Q}_n(G) = \begin{cases}
\frac1{nE(E-1)\dotsm (E-(k-1))}\text{,}&\text{if } E\ge k\\
0\text{,}& \text{otherwise}
\end{cases}
\end{equation*}
where $E=nL'(1)$ is the number of edges in the whole Tanner graph, 
as defined in Subsection~\ref{subsec:irregens}, 
and where $k$ denotes the number of edges in $G$.
For convenience, we will use a marginalized probability $\Pn(\cdot)$ 
which is induced from $\mathbb{Q}_n(\cdot)$ via the equivalence relation defined as follows: 
$G=(\mathcal{V_N}, \mathcal{C_N}, v_0, \mathcal{S_{V_N}}, \mathcal{S_{C_N}}, \mathcal{E_N})$ and
$G'=(\mathcal{V_N}', \mathcal{C_N}', v_0', \mathcal{S_{V_N'}}, \mathcal{S_{C_N'}}, \mathcal{E_N}')$ are
equivalent if and only if there exist bijections
$\sigma_V: \mathcal{S_{V_N}} \to \mathcal{S_{V_N'}}$ and
$\sigma_C: \mathcal{S_{C_N}} \to \mathcal{S_{C_N'}}$
such that
\begin{enumerate}
\item[c1.] $(\sigma_V(s),\sigma_C(t))\in\mathcal{E_N}'$ for all $(s,t)\in\mathcal{E_N}$
\item[c2.] $\sigma_V(v_0)=v_0'$
\item[c3.] $\forall v\in\mathcal{V_N}\backslash v_0,
\exists v'\in\mathcal{V_N}'\backslash v_0'$ s.t. $\sigma_V(v) \overset{c}{=} v'$
\item[c4.] $\forall c\in\mathcal{C_N},
\exists c'\in\mathcal{C_N}'$ s.t. $\sigma_C(c) \overset{c}{=} c'$
\end{enumerate}
where $\sigma_V(v)$ (respectively $\sigma_C(c)$) are tuples whose $i$-th element is the image of $i$-th element of $v$ (respectively $c$) under $\sigma_V$ (respectively $\sigma_C$)
and where $m\overset{c}{=}m'$ if and only if they are equal under cyclic shift for tuples $m$ and $m'$ of sockets.
This equivalence relation is weaker than what is used in $\mathbb{Q}_n(\cdot)$ 
and stronger than the conventional equivalence relation in graph theory 
which does not distinguish sockets.

Under this equivalence relation, the number of neighborhood graphs equivalent to $G$ is
\begin{multline*}
nL_{|v_0|}\left(\prod_i\prod_{l=0}^{v_i-1} (nL_i-l)i\right)\left( \prod_j\prod_{l=0}^{c_j-1} (mR_j-l)j\right)\\
=nL_{|v_0|}\left(\prod_i\prod_{l=0}^{v_i-1} (E\lambda_i-li)\right)\left(\prod_j\prod_{l=0}^{c_j-1} (E\rho_j-lj)\right)
\end{multline*}
where $v_i$ denotes the number of variable nodes of degree $i$ in $G$, where $c_j$ denotes the number of check nodes of degree $j$ in $G$, 
and where $m$ denotes the number of check nodes in the whole Tanner graph, i.e.,\ $m=nL'(1)/R'(1)$.
Hence, the probability $\Pn(\cdot)$ which marginalizes equivalent neighborhood graphs
is given as
\begin{multline}
\Pn(G)=
L_{|v_0|}\frac{\prod_i\prod_{l=0}^{v_i-1} (E\lambda_i-li) \prod_j\prod_{l=0}^{c_j-1} (E\rho_j-lj)}{\prod_{i=0}^{k-1}(E-i)}\text{.}\label{probn}
\end{multline}
This defines the probability associated with a neighborhood graph $G$ 
in the neighborhood graph ensemble.  

Since $E=\Theta(n)$, the denominator and the numerator are $\Theta(n^k)$ and $\Theta(n^w)$, respectively,
where $w$ denotes the number of nodes in $G$ except the root node.
One therefore has $\Pn(G)=\Theta(n^{w-k})$.
Since the number of cycles in $G$ is $(k-w)$, the next lemma follows.

\begin{lemma}\label{kcycles}
For a neighborhood graph $G$ which has $c$ cycles,
\begin{equation*}
\Pn(G) = \Theta(n^{-c})\text{.}
\end{equation*}
\end{lemma}
\noindent
This lemma plays a key role in this paper.
Classification of neighborhood graphs according to the number of cycles is also considered in~\cite{Mon06}.



\subsection{Tree ensembles}
From Lemma~\ref{kcycles}, neighborhood graphs of a fixed depth with cycles are not generated in the large-blocklength limit.
To be precise,
\begin{equation}
\mathbb{P}_\infty(G) := \lim_{n\to\infty}\Pn(G) =
L_{|v_0|}\prod_{v\in \mathcal{V}(G)\backslash v_0} \lambda_{|v|} \prod_{c\in \mathcal{C}(G)} \rho_{|c|}\label{probtree}
\end{equation}
for a tree graph $G$ and $\mathbb{P}_\infty(G) = 0$ for any graph $G$ with cycles.
The ensemble of tree neighborhood graphs with probability $\mathbb{P}_\infty(G)$ is called the \textit{tree ensemble from node perspective}, 
and is denoted by $\mathring{\mathcal{T}}_t(\lambda(x),\rho(x))$.

We also define two other tree neighborhood graph ensembles, namely 
tree neighborhood graph ensembles from edge perspective
$\vec{\mathcal{T}}_t^v(\lambda(x),\rho(x))$ and $\vec{\mathcal{T}}_t^c(\lambda(x),\rho(x))$.
Neighborhood graphs in $\vec{\mathcal{T}}_t^v(\lambda(x),\rho(x))$ and $\vec{\mathcal{T}}_t^c(\lambda(x),\rho(x))$ are rooted at
an edge incident to a variable node and a check node, respectively.
The number of check nodes in the shortest path from the node connected to the root edge to any node is not greater than $t$. 
Only variable nodes which have distance $t$ from the node connected to the root edge have sockets which do not connect to any edges. 
The probability of a neighborhood graph $G$ rooted at an edge in both ensembles is
\begin{equation*}
\prod_{v\in \mathcal{V}(G)} \lambda_{|v|} \prod_{c\in \mathcal{C}(G)} \rho_{|c|}\text{.}
\end{equation*}
The ensembles $\mathring{\mathcal{T}}_t(\lambda(x),\rho(x))$ and $\vec{\mathcal{T}}_t^v(\lambda(x),\rho(x))$ are also defined in~\cite{RiU05/LTHC}.

\section{Main result}
\subsection{The decomposition of the coefficient of $n^{-1}$}\label{subsec:dec}
For each variable node, an error occurrence after $t$ BP iterations depends only on a realization of a neighborhood graph $G$ of depth $t$ and
realizations of channel outputs corresponding to variable nodes in $G$.
In other words, the bit error probability of irregular ensemble is
\begin{equation}
\Pb(n,\epsilon,t) = \sum_{G\in\mathcal{G}_t} \Pn(G)\Pb(\epsilon,G)
\end{equation}
where $\mathcal{G}_t$ denotes the set of all neighborhood graphs of depth $t$,
and where 
$\Pb(\epsilon,G)$ denotes the error probability of the root node of $G$ after $t$ iterations
when the erasure probability of each node in $G$ is initialized with $\epsilon$.
From Lemma~\ref{kcycles}, it holds that
\begin{equation*}
\Pb(\infty,\epsilon,t)
= \sum_{G\in\mathcal{T}_t} \mathbb{P}_\infty(G)\Pb(\epsilon, G) 
\end{equation*}
where $\mathcal{T}_t$ denotes the set of all cycle-free neighborhood graphs.
This fact allows us to calculate the limit of the bit error probability $\Pb(\infty,\epsilon,t) := \lim_{n\to\infty}\Pb(n,\epsilon,t)$ in a recursive manner, leading to the idea of density evolution.
\begin{lemma}[Density evolution~\cite{910577}]\label{lem:debec}
Let $Q_\epsilon(t)$ denote erasure probability of messages into check nodes at $t$-th iteration, 
and let
$P_\epsilon(t)$ denote erasure probability of messages into variable nodes at $t$-th iteration in the limit of infinite blocklength.
Then
\begin{align*}
\Pb(\infty,\epsilon,t) &= \epsilon L(P_\epsilon(t))\\
Q_\epsilon(t) &= \epsilon\lambda(P_\epsilon(t-1))\\
P_\epsilon(t) &=
\begin{cases}
1,&\text{if } t = 0\\
1-\rho(1-Q_\epsilon(t)),&\text{otherwise.}
\end{cases}
\end{align*}
\end{lemma}

On the other hand, one observes from Lemma~\ref{kcycles} that the second and the third dominant terms are $\Theta(n^{-1})$ and $\Theta(n^{-2})$, respectively.
In other words, one has the following large-$n$ asymptotic expansion of $\Pb(n,\epsilon,t)$: 
\begin{equation*}
\Pb(n,\epsilon,t)=\Pb(\infty,\epsilon,t) + \alpha(\epsilon,t)\frac1{n} + \Theta\left(\frac1{n^2}\right)
\end{equation*}
where the coefficient $\alpha(\epsilon,t)$ of $n^{-1}$ is defined as
\begin{equation*}
\alpha(\epsilon,t) := \lim_{n\to\infty} n(\Pb(n,\epsilon,t) - \Pb(\infty,\epsilon,t)) \text{.}
\end{equation*}
Moreover, Lemma~\ref{kcycles} tells us that 
$\alpha(\epsilon,t)$ can be decomposed into two components as follows:
\begin{align*}
\alpha(\epsilon,t) &= \lim_{n\to\infty} n\left(\sum_{G\in\mathcal{T}_t} \Pn(G)\Pb(\epsilon,G) - \Pb(\infty,\epsilon,t)\right)\nonumber\\
& \hspace{9em} +\lim_{n\to\infty} n\sum_{G\in\mathcal{S}_t} \Pn(G)\Pb(\epsilon,G)\nonumber\\
&=: \beta(\epsilon,t)+\gamma(\epsilon,t) 
\end{align*}
where $\mathcal{S}_t$ denotes the set of all single-cycle neighborhood graphs 
and where the components $\beta(\epsilon,t)$ and $\gamma(\epsilon,t)$ represent contributions 
of cycle-free and single-cycle neighborhood graphs, respectively.  
In Subsection~\ref{subsec:beta} and Subsection~\ref{subsec:gamma},
recursive formulas to evaluate 
$\beta(\epsilon,t)$ and $\gamma(\epsilon,t)$ for $(\lambda(x),\rho(x))$-irregular ensembles are derived, respectively.

\subsection{The contribution of cycle-free neighborhood graphs}\label{subsec:beta}
The component $\beta(\epsilon,t)$ for cycle-free neighborhood graphs is calculated as
\begin{align*}
\beta(\epsilon,t)&:=\lim_{n\to\infty} n\left(\sum_{G\in\mathcal{T}_t} \Pn(G)\Pb(\epsilon,G) - \Pb(\infty,\epsilon,t)\right)\\
&=\sum_{G\in\mathcal{T}_t} \left[\lim_{n\to\infty} n\left(\Pn(G)-\mathbb{P}_\infty(G)\right)\right]\Pb(\epsilon,G)\text{.}
\end{align*}

From (\ref{probn}) and (\ref{probtree}), the contributions of a cycle-free neighborhood graph $G$ to $\beta(\epsilon,t)$ is obtained as
\begin{multline*}
 L_{|v_0|}\prod_{v\in \mathcal{V}(G)\backslash v_0} \lambda_{|v|} \prod_{c\in \mathcal{C}(G)} \rho_{|c|} \Pb(\epsilon,G)\\
\lim_{n\to\infty}n\left(\frac{\prod_i \prod_{l=0}^{v_i-1}\left(E-l\frac{i}{\lambda_i}\right) \prod_j \prod_{l=0}^{c_j-1}\left(E-l\frac{j}{\rho_j}\right)}{\prod_{i=0}^{k-1}(E-i)}-1\right)\\
= L_{|v_0|}\prod_{v\in \mathcal{V}(G)\backslash v_0} \lambda_{|v|} \prod_{c\in \mathcal{C}(G)} \rho_{|c|} \Pb(\epsilon,G)\frac1{2L'(1)}\\
\times\left(k(k-1)-\sum_i\frac{i}{\lambda_i}v_i(v_i-1)-\sum_j\frac{j}{\rho_j}c_j(c_j-1) \right)
\text{.}
\end{multline*}
Hence, $\beta(\epsilon,t)$ is obtained via expectation, denoted by $\mathbb{E}_t[\cdot]$, 
on the tree ensemble $\mathring{\mathcal{T}}_t(\lambda(x),\rho(x))$ 
of depth $t$ from node perspective as 
\begin{multline}
\frac1{2L'(1)}\bigg[\mathbb{E}_t[K(K-1)P] - \sum_i\frac{i}{\lambda_i}\mathbb{E}_t[V_i(V_i-1)P]\\
- \sum_j\frac{j}{\rho_j}\mathbb{E}_t[C_j(C_j-1)P]\bigg]\label{eqbeta}
\end{multline}
where $K$, $V_i$ and $C_j$ denote random variables representing the number of edges,
the number of variable nodes of degree $i$, 
and the number of check nodes of degree $j$, respectively, 
and where $P$ denotes the erasure probability of the root node after $t$ BP iterations.

The three expectations in~\eqref{eqbeta} are obtained using generating functions as
\begin{align}
\mathbb{E}_t[K(K-1)P] &= \left.\frac{\partial^2 \mathbb{E}_t[x^KP]}{\partial x^2}\right|_{x=1}\label{genk}\\
\mathbb{E}_t[V_i(V_i-1)P] &= \left.\frac{\partial^2 \mathbb{E}_t[x^{V_i}P]}{\partial x^2}\right|_{x=1}\label{genv}\\
\mathbb{E}_t[C_j(C_j-1)P] &= \left.\frac{\partial^2 \mathbb{E}_t[x^{C_j}P]}{\partial x^2}\right|_{x=1}\text{.}\label{genc}
\end{align}
In order to deal with these generating functions, 
we now define the following ``canonical'' generating function: 
\begin{equation}
\Phi(t;\,\{y_k\},\,\{z_l\})
=\mathbb{E}_t\left[\prod_ky_k^{V_k}\prod_lz_l^{C_l}P\right].
\label{eq:can}
\end{equation}
The three generating functions that appear in 
the right-hand sides of~\eqref{genk}--\eqref{genc} are obtained from $\Phi(t;\,\{y_k\},\,\{z_l\})$ as 
\begin{align*}
\mathbb{E}_t[x^KP] &= \frac1{x}\left.
\Phi(t;\,\{y_k\},\,\{z_l\})
\right|_{y_k=x,z_l=x\text{ for all } k,l}\\
\mathbb{E}_t[x^{V_i}P] &= \left.
\Phi(t;\,\{y_k\},\,\{z_l\})
\right|_{y_i=x;\;y_k=1,\,{}^\forall k\ne i;\; z_l=1,\,{}^\forall l}\\
\mathbb{E}_t[x^{C_j}P] &= \left.
\Phi(t;\,\{y_k\},\,\{z_l\})
\right|_{z_j=x;\;y_k=1,\,{}^\forall k;\; z_l=1, {}^\forall l\ne j}\text{.}
\end{align*}
The key idea here is that 
one can evaluate the canonical generating function $\Phi(t;\,\{y_k\},\,\{z_l\})$ 
via extending density evolution in such a way that 
``densities'' to be updated in density evolution incorporate 
the auxiliary variables $\{y_k\}$ and $\{z_l\}$.  
We call our extension the augmented density evolution.  
In the conventional density evolution,
expectation of density of messages over a tree ensemble is calculated in a recursive way.
In the augmented density evolution, on the other hand, 
one considers, for each tree, a product of the density of messages at the root node and a monomial 
reflecting degree histogram of the tree, 
and calculate its expectation over the tree ensemble, 
which can be performed recursively in a similar way to density evolution. 
The canonical generating function $\Phi(t;\,\{y_k\},\,\{z_l\})$ in the general case is thus 
a polynomial whose coefficients are conical combinations of densities. 
Since we are assuming $\mathrm{BEC}(\epsilon)$, 
we only have to deal with erasure probabilities of messages instead of densities of messages, 
as shown in Lemma~\ref{lem:debec}.
Hence, the canonical generating function $\Phi(t;\,\{y_k\},\,\{z_l\})$ 
is obtained by a recursive calculation of polynomials in $\{y_k\}$ and $\{z_l\}$ 
with real-valued coefficients.
The next lemma provides a set of recursive formulas 
to evaluate the canonical generating function $\Phi(t;\,\{y_k\},\,\{z_l\})$.  

\begin{lemma}\label{gen}
The canonical generating function $\Phi(t;\,\{y_k\},\,\{z_l\})$ is 
given by 
\begin{equation*}
\Phi(t;\;\{y_k\},\,\{z_l\}) = \epsilon \mathfrak{L}(F(t))\\
\end{equation*}
where
\begin{align*}
F(t) &:= \begin{cases}
1,&\text{if } t = 0\\
\mathcal{P}(g(t)) - \mathcal{P}(G(t)),& \text{otherwise}\\
\end{cases}\\
G(t) &:=
\mathcal{L}(f(t-1)) - \epsilon \mathcal{L}(F(t-1))\\
f(t) &:= \begin{cases}
1,&\text{if }t=0\\
\mathcal{P}(g(t)),&\text{otherwise}
\end{cases}\\
g(t) &:= \mathcal{L}(f(t-1))
\end{align*}
and where
\begin{align*}
\mathfrak{L}(x) &:= \sum_i L_i y_i x^i \\
\mathcal{L}(x) &:= \sum_i \lambda_i y_i x^{i-1} \\
\mathcal{P}(x) &:= \sum_j \rho_j z_j x^{j-1} \text{.}
\end{align*}
\end{lemma}
\begin{IEEEproof}
The generating function is calculated as
\begin{align*}
&\mathbb{E}_t\left[\prod_ky_k^{V_k}\prod_lz_l^{C_l}P\right]\\
&\quad=\mathbf{E}_m\left[y_m \epsilon \left(\mathbb{E}_t^c\left[\prod_ky_k^{V_k}\prod_lz_l^{C_l}Q\right]\right)^{m}\right]\\
&\quad=\epsilon \mathfrak{L}\left(\mathbb{E}_t^c\left[\prod_ky_k^{V_k}\prod_lz_l^{C_l}Q\right]\right)
\end{align*}
where $m$ denotes a random variable corresponding to the degree of the root node,
where $\mathbf{E}_m$ denotes expectation with respect to the degree of the root node,
where $\mathbb{E}_t^c[\cdot]$ denotes expectation on $\vec{\mathcal{T}}_t^c$, and
where $Q$ denotes a random variable corresponding to erasure probability of messages transmitted to the root edge at $t$-th iteration.
Now define
\begin{align*}
f(t)&= \mathbb{E}_t^c\left[\prod_ky_k^{V_k}\prod_lz_l^{C_l}\right]\\
g(t)&= \mathbb{E}_{t-1}^v\left[\prod_ky_k^{V_k}\prod_lz_l^{C_l}\right]\\
F(t)&= \mathbb{E}_t^c\left[\prod_ky_k^{V_k}\prod_lz_l^{C_l}Q\right]\\
G(t)&= \mathbb{E}_{t-1}^v\left[\prod_ky_k^{V_k}\prod_lz_l^{C_l}(1-P)\right]
\end{align*}
where $\mathbb{E}_t^v[\cdot]$ denotes expectation on $\vec{\mathcal{T}}_t^v$.
The functions $f(t)$ and $g(t)$ are the generating functions of $\{V_k\}$ and $\{C_l\}$ 
on the ensembles $\vec{\mathcal{T}}_t^c$ and $\vec{\mathcal{T}}_{t-1}^v$, respectively.  
The functions $F(t)$ and $G(t)$ are
reweighted versions of the generating functions, where reweighting is done on the basis of 
erasure probability at the root node.
It should be noted that dependence of these functions on the auxiliary variables $\{y_k\}$ and $\{z_l\}$ 
is implicit in the notation.  
The desired expectations are calculated recursively as
\begin{align*}
f(0)&=F(0)=1\\
f(t)
&=  \mathbb{E}_t^c\left[z_m g(t)^{m-1}\right]=\mathcal{P}(g(t))\text{,}&\text{if } t\ge 1\\
g(t)
&= \mathbb{E}_{t-1}^v\left[y_m f(t-1)^{m-1}\right]=\mathcal{L}(f(t-1)) 
\end{align*}
\begin{align*}
F(t)
&=  \mathbb{E}_t^c\left[z_m\left(g(t)^{m-1}-G(t)^{m-1}\right)\right]\\
&= f(t) - \mathcal{P}(G(t))\text{,}& \text{if } t\ge 1 \\
G(t)
&=  \mathbb{E}_{t-1}^v\left[y_m\left(f(t-1)^{m-1}-\epsilon F(t-1)^{m-1}\right)\right]\\
&= g(t) - \epsilon\mathcal{L}(F(t-1)).
\end{align*}
\end{IEEEproof}

\begin{figure*}[!t]
\psfrag{t1}{$s_1=1$, $s_2=6$}
\psfrag{t2}{$s_1=1$, $s_2=5$}
\psfrag{t3}{$s_1=0$, $s_2=4$}
\psfrag{t4}{$s_1=0$, $s_2=5$}
\psfrag{t5}{$s=6$}
\psfrag{t6}{$s=5$}
\psfrag{u1}{Type I}
\psfrag{u2}{Type II}
\psfrag{u3}{Type III}
\psfrag{u4}{Type IV}
\psfrag{u5}{Type V}
\psfrag{u6}{Type VI}
\includegraphics[width=\hsize]{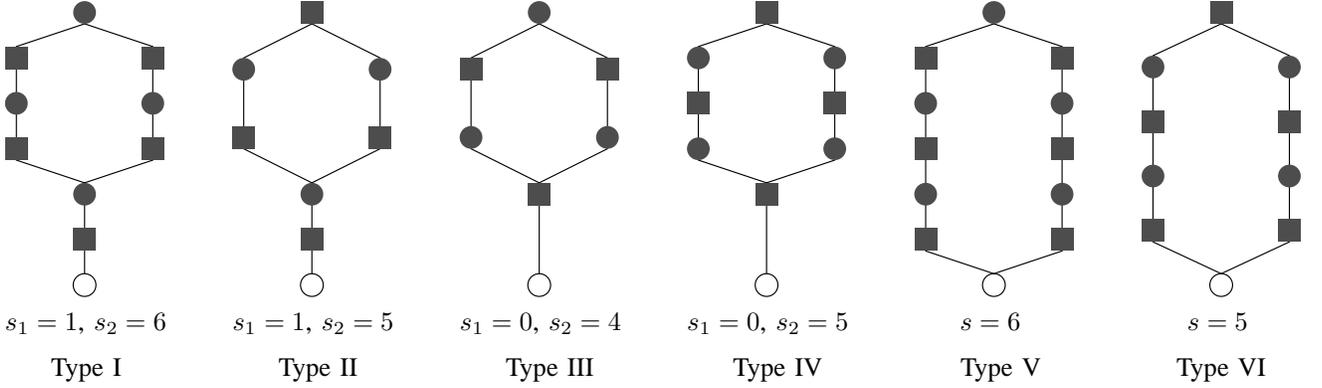}
\caption{Six types of single-cycle neighborhood graphs.
All nodes which are not included in the two minimum path from the root node to the deepest node in the cycle are not described in the above figure.
These are classified according to whether the shallowest and the deepest nodes in the cycle are variable nodes, check nodes or the root node.
A depth of the shallowest node in the cycle corresponds to $s_1$.
The number of nodes in the shortest path from the root node to the deepest node in the cycle corresponds to $s_2+1$ and $s+1$.}
\label{sc}
\end{figure*}

Considering appropriate derivatives of the recursive formulas 
given by Lemma~\ref{gen}, one obtains explicit formulas 
to evaluate the three expectations in~\eqref{eqbeta} recursively, 
on the basis of which one can evaluate $\beta(\epsilon,t)$ explicitly.  
The derivation is elaborate but straightforward, 
so that we omit details of the derivation and only show the end result.  
Let us define, for $n=1$ and $2$, 
\begin{align*}
f^{(n)}(t)&:=\left.\frac{\partial^n\!\! \left.f(t)\right|_{y_k=x,z_l=x\text{ for all }k,l}}{\partial x^n}\right|_{x=1}\\
f_v^{(n)}(t,i)&:=\left.\frac{\partial^nf(t)}{\partial y_i^n}\right|_{y_k=1,\,z_l=1\text{ for all }k,l}\\
f_c^{(n)}(t,j)&:=\left.\frac{\partial^nf(t)}{\partial z_j^n}\right|_{y_k=1,\,z_l=1\text{ for all }k,l}.
\end{align*}
Similar definitions are applied to $g(t)$, $F(t)$ and $G(t)$ 
to define $g^{(n)}(t)$, $g_v^{(n)}(t,i)$, etc.  
The resulting 24 functions are to be used to evaluate 
the relevant expectations, and the recursive formulas 
of these functions used in the evaluation 
are summarized in the next theorem.  

\setcounter{equation}{6}
\begin{theorem}\label{beta}
$\beta(\epsilon,t)$ for $(\lambda(x),\rho(x))$-irregular ensembles is calculated as
\begin{multline*}
\beta(\epsilon,t) = \frac1{2L'(1)}\bigg[\mathbb{E}_t[K(K-1)P]\\
-\sum_i\frac{i}{\lambda_i}\mathbb{E}_t[V_i(V_i-1)P]-\sum_j\frac{j}{\rho_j}\mathbb{E}_t[C_j(C_j-1)P]\bigg]
\end{multline*}
where $\mathbb{E}_t[K(K-1)P]$, $\mathbb{E}_t[V_i(V_i-1)P]$ and $\mathbb{E}_t[C_j(C_j-1)P]$ are calculated by (\ref{bk}), (\ref{bv}) and (\ref{bc}), respectively.
The functions $P_\epsilon(t)$ and $Q_\epsilon(t)$ appearing in these formulas 
are to be evaluated recursively via the conventional density evolution (Lemma 2).

\begin{align*}
f^{(1)}(t) &= \begin{cases}
0\text{,}&\text{if }t=0\\
1+\rho'(1)g^{(1)}(t)\text{,}&\text{otherwise}
\end{cases}\\
g^{(1)}(t) &= 1+\lambda'(1)f^{(1)}(t-1)\\
F^{(1)}(t) &=\begin{cases}
0\text{,}&\text{if } t = 0\\
f^{(1)}(t) - \rho(1-Q_\epsilon(t))\\
\quad - \rho'(1-Q_\epsilon(t))G^{(1)}(t)\text{,}&\text{otherwise}\\
\end{cases}\\
G^{(1)}(t) &= g^{(1)}(t) - \epsilon \lambda(P_\epsilon(t-1))\\
&\quad - \epsilon \lambda'(P_\epsilon(t-1))F^{(1)}(t-1)
\end{align*}
\begin{align*}
f^{(2)}(t) &= \begin{cases}
0\text{,}&\text{if }t=0\\
2\rho'(1)g^{(1)}(t)\\
\quad +\rho''(1)g^{(1)}(t)^2+\rho'(1)g^{(2)}(t)\text{,}&\text{otherwise}
\end{cases}\\
g^{(2)}(t) &= 2\lambda'(1)f^{(1)}(t-1)+\lambda''(1)f^{(1)}(t-1)^2\\
&\quad +\lambda'(1)f^{(2)}(t-1)\\
F^{(2)}(t) &= \begin{cases}
0\text{,}&\text{if } t = 0\\
f^{(2)}(t) - 2\rho'(1-Q_\epsilon(t))G^{(1)}(t)\\
\quad - \rho''(1-Q_\epsilon(t))G^{(1)}(t)^2\\
\quad - \rho'(1-Q_\epsilon(t))G^{(2)}(t)\text{,}&\text{otherwise}\\
\end{cases}\\
G^{(2)}(t) &=
g^{(2)}(t) - 2\epsilon\lambda'(P_\epsilon(t-1))F^{(1)}(t-1)\\
&\quad - \epsilon\lambda''(P_\epsilon(t-1))F^{(1)}(t-1)^2\\
&\quad - \epsilon\lambda'(P_\epsilon(t-1))F^{(2)}(t-1)
\end{align*}
\begin{equation}
\mathbb{E}_t[K(K-1)P] = \epsilon L''(P_\epsilon(t))F^{(1)}(t)^2 + \epsilon L'(P_\epsilon(t))F^{(2)}(t)\label{bk}
\end{equation}
\begin{align*}
f_v^{(1)}(t,i) &= \begin{cases}
0\text{,}&\text{if }t=0\\
\rho'(1)g_v^{(1)}(t,i)\text{,}&\text{otherwise}
\end{cases}\\
g_v^{(1)}(t,i) &= \lambda'(1)f_v^{(1)}(t-1,i) + \lambda_i\\
F_v^{(1)}(t,i) &=\begin{cases}
0\text{,}&\text{if } t = 0\\
f_v^{(1)}(t,i) - \rho'(1-Q_\epsilon(t))G_v^{(1)}(t,i)\text{,}&\text{otherwise}\\
\end{cases}\\
G_v^{(1)}(t,i) &= g_v^{(1)}(t,i) - \epsilon \lambda'(P_\epsilon(t-1))F_v^{(1)}(t-1,i)\\
&\quad - \epsilon \lambda_i P_\epsilon(t-1)^{i-1}
\end{align*}

\begin{align*}
f_v^{(2)}(t,i) &= \begin{cases}
0\text{,}&\text{if }t=0\\
\rho''(1)g_v^{(1)}(t,i)^2 + \rho'(1)g_v^{(2)}(t,i)\text{,}&\text{otherwise}
\end{cases}\\
g_v^{(2)}(t,i) &= \lambda''(1)f_v^{(1)}(t-1,i)^2 + \lambda'(1)f_v^{(2)}(t-1,i)\\
&\quad + 2\lambda_i(i-1)f_v^{(1)}(t-1,i)\\
F_v^{(2)}(t,i) &= \begin{cases}
0\text{,}&\text{if } t = 0\\
f_v^{(2)}(t,i) - \rho''(1-Q_\epsilon(t)) G_v^{(1)}(t,i)^2\\
\quad - \rho'(1-Q_\epsilon(t)) G_v^{(2)}(t,i)\text{,}&\text{otherwise}
\end{cases}\\
G_v^{(2)}(t,i) &=
g_v^{(2)}(t,i) -\epsilon\lambda''(P_\epsilon(t-1))F_v^{(1)}(t-1,i)^2\\
&\quad - \epsilon\lambda'(P_\epsilon(t-1))F_v^{(2)}(t-1,i)\\
&\quad - 2\epsilon\lambda_i(i-1)P_\epsilon(t-1)^{i-2}F_v^{(1)}(t-1,i)
\end{align*}
\begin{multline}
\mathbb{E}_t[V_i(V_i-1)P] = \epsilon L''(P_\epsilon(t)) F_v^{(1)}(t,i)^2 \\
+ \epsilon L'(P_\epsilon(t)) F_v^{(2)}(t,i)
+ 2 \epsilon L_i i P_\epsilon(t)^{i-1} F_v^{(1)}(t,i)\label{bv}
\end{multline}
\begin{align*}
f_c^{(1)}(t,j) &= \begin{cases}
0\text{,}&\text{if }t=0\\
\rho'(1)g_c^{(1)}(t,j) + \rho_j\text{,}&\text{otherwise}
\end{cases}\\
g_c^{(1)}(t,j) &= \lambda'(1)f_c^{(1)}(t-1,j)\\
F_c^{(1)}(t,j) &= \begin{cases}
0\text{,}&\text{if } t = 0\\
f_c^{(1)}(t,j) - \rho'(1-Q_\epsilon(t))G_c^{(1)}(t,j)\\
\quad - \rho_j (1-Q_\epsilon(t))^{j-1}\text{,}& \text{otherwise}\\
\end{cases}\\
G_c^{(1)}(t,j) &=
g_c^{(1)}(t,j) - \epsilon \lambda'(P_\epsilon(t-1)) F_c^{(1)}(t-1,j)
\end{align*}

\begin{align*}
f_c^{(2)}(t,j) &= \begin{cases}
0\text{,}&\text{if }t=0\\
\rho''(1)g_c^{(1)}(t,j)^2 + \rho'(1)g_c^{(2)}(t,j)\\
\quad + 2\rho_j(j-1)g_c^{(1)}(t,j)\text{,}&\text{otherwise}
\end{cases}\\
g_c^{(2)}(t,j) &= \lambda''(1)f_c^{(1)}(t-1,j)^2 + \lambda'(1)f_c^{(2)}(t-1,j)\\
F_c^{(2)}(t,j) &= \begin{cases}
0\text{,}&\text{if } t = 0\\
f_c^{(2)}(t,j) - \rho''(1-Q_\epsilon(t))G_c^{(1)}(t,j)^2\\
\quad - \rho'(1-Q_\epsilon(t))G_c^{(2)}(t,j)\\
\quad - 2\rho_j(j-1)(1-Q_\epsilon(t))^{j-2}\\
\quad \times G_c^{(1)}(t,j)\text{,}& \text{otherwise}\\
\end{cases}\\
G_c^{(2)}(t,j) &=
g_c^{(2)}(t,j) - \epsilon \lambda''(P_\epsilon(t-1))F_c^{(1)}(t-1,j)^2\\
&\quad - \epsilon \lambda'(P_\epsilon(t-1)) F_c^{(2)}(t-1,j)
\end{align*}
\begin{equation}
\mathbb{E}_t[C_j(C_j-1)P] = \epsilon L''(P_\epsilon(t)) F_c^{(1)}(t,j)^2 + \epsilon L'(P_\epsilon(t)) F_c^{(2)}(t,j)\text{.}\label{bc}
\end{equation}
\end{theorem}



\subsection{Method of enumeration}\label{subsec:coef}
In order to calculate the coefficient $\alpha(\epsilon ,t)$ of $n^{-1}$,
it is necessary to evaluate the contribution
of single-cycle neighborhood graphs, i.e.,
\begin{equation*}
\gamma(\epsilon,t) := \lim_{n\to\infty} n\sum_{G\in\mathcal{S}_t} \Pn(G)\Pb(\epsilon,G)\text{.} 
\end{equation*}
For ease of the explanation of how to evaluate $\gamma(\epsilon,t)$, 
which is deferred to Subsection~\ref{subsec:gamma}, 
we consider in this subsection a different quantity, namely
the coefficient of $n^{-1}$ in the probability of single-cycle neighborhood graphs: 
\begin{equation*}
\xi(t) := \lim_{n\to\infty} n\sum_{G\in\mathcal{S}_t} \Pn(G)\text{.} 
\end{equation*}
Methods for enumeration of $\xi(t)$ introduced in this subsection will be 
extended to those for calculation of $\gamma(\epsilon,t)$ in Subsection~\ref{subsec:gamma}.
In both calculations, we consider subgraph $S(G)$ of a single-cycle neighborhood graph $G$
consisting of nodes which are included by the two shortest paths from the root node to the deepest node in the cycle.
We classify single-cycle neighborhood graphs into six types of subgraphs $S(G)$ as shown in Fig.~\ref{sc}.
They are classified according to whether the shallowest node in the cycle is a non-root variable, a check, or the root node, 
as well as whether the deepest node in the cycle is a variable or check node.
Types I to IV of neighborhood graphs have two parameters: $s_1$ corresponding to the depth of the shallowest node in the cycle, and
$s_2$ for which $s_2+1$ equals to the number of nodes in the shortest path from the root node to the deepest node in the cycle.
Types V and VI of neighborhood graphs have a parameter $s$ which plays the same role as $s_2$ in Types I to IV.
The set of single-cycle neighborhood graphs of Type I and Type II with the parameters $s_1$ and $s_2$ is denoted by $\mathcal{S}_v(t,s_1,s_2)$.
The sets $\mathcal{S}_c(t,s_1,s_2)$ and $\mathcal{S}_r(t,s)$ are defined in the similar way.

We consider marginalization of the probability using the classification of neighborhood graphs.
The probability $\Pn(G)$ of a single-cycle neighborhood graph $G$ is
\begin{equation*}
L_{|v_0|}\frac{\prod_i \prod_{l=0}^{v_i-1}(\lambda_iE-li) \prod_j \prod_{l=0}^{c_j-1}(\rho_jE-lj)}{\prod_{i=0}^{k-1}(E-i)}\text{.}
\end{equation*}
Since $E=nL'(1)$, we obtain the coefficient of $n^{-1}$ as
\begin{equation*}
\lim_{n\to\infty}n\Pn(G)=\frac1{L'(1)}L_{|v_0|}\prod_{v\in \mathcal{V}(G)}\lambda_{|v|} \prod_{c\in \mathcal{C}(G)} \rho_{|c|}\text{.}
\end{equation*}
In order to enumerate the coefficient of $n^{-1}$ in the probability of single-cycle neighborhood graphs,
we consider an equivalence relation 
in which positions of sockets connected to a socket associated with a node in $S(G)$ are not distinguished, 
which is weaker than what is used in $\Pn(\cdot)$.  
The sets of representatives of the resulting equivalence classes in $\mathcal{S}_v(t,s_1,s_2)$, $\mathcal{S}_c(t,s_1,s_2)$ and $\mathcal{S}_r(t,s)$
are denoted by $\bar{\mathcal{S}}_v(t,s_1,s_2)$, $\bar{\mathcal{S}}_c(t,s_1,s_2)$ and $\bar{\mathcal{S}}_r(t,s)$, respectively.
The coefficients of $n^{-1}$ in the probability of 
single-cycle neighborhood graphs of Type I and Type II with parameters $s_1$ and $s_2$ 
are evaluated in a unified way ($s_2$ is even for Type I and odd for Type II), 
and are obtained as
\begin{multline}
\lim_{n\to\infty}\sum_{G\in \mathcal{S}_{v}(t,s_1,s_2)}n\Pn(G)
=\frac1{L'(1)} \sum_{G\in \bar{\mathcal{S}}_{v}(t,s_1,s_2)}\\
\quad L_{|v_0|} |v_0|\prod_{v\in \mathcal{V}(S(G))\backslash \{v_0,w\}}\lambda_{|v|}(|v|-1) \prod_{c\in \mathcal{C}(S(G))} \rho_{|c|}(|c|-1)\\
\quad \times\lambda_{|w|}\binom{|w|-1}{2}\prod_{v\in \mathcal{V}(G)\backslash \mathcal{V}(S(G))}\lambda_{|v|}
\prod_{c\in \mathcal{C}(G)\backslash \mathcal{C}(S(G))} \rho_{|c|}\\
= \frac1{2}\lambda''(1)\rho'(1)^2(\lambda'(1)\rho'(1))^{s_2-s_1-2} \label{eq:enumcoef}
\end{multline}
where $w$ denotes the shallowest variable node in the cycle.
In the first equality in~\eqref{eq:enumcoef}, single-cycle neighborhood graphs of Type I or Type II are marginalized
according to the equivalence relation.
In the second equality, by the marginalizations, quantities corresponding to nodes not included in $S(G)$ become $1$,
and quantities corresponding to the root node, the shallowest node in the cycle, other variable nodes in $S(G)$, and check nodes in $S(G)$
become $L'(1)$, $\lambda''(1)/2$, $\lambda'(1)$ and $\rho'(1)$, respectively.
The concept of the equivalence classes $\bar{\mathcal{S}}_v(t,s_1,s_2)$, $\bar{\mathcal{S}}_c(t,s_1,s_2)$ and $\bar{\mathcal{S}}_r(t,s)$
is useful not only for the calculation~\eqref{eq:enumcoef} but also for the calculation of $\gamma(\epsilon,t)$ in Subsection~\ref{subsec:gamma}.

In the same way, the coefficients of $n^{-1}$ in the probability of single-cycle neighborhood graphs of Type III and Type IV 
with parameters $s_1$ and $s_2$ are calculated as
\begin{equation*}
\frac1{2}\rho''(1)\lambda'(1)(\lambda'(1)\rho'(1))^{s_2-s_1-2}
\end{equation*}
and those for Type V and Type VI with the parameter $s$ are calculated as
\begin{equation*}
\frac1{2}(\lambda'(1)\rho'(1))^s\text{.}
\end{equation*}
Similar calculations are also used in~\cite{Mon06}.
The classification of single-cycle neighborhood graphs in this subsection is
finer than that in~\cite{Mon06} for the purpose of calculation of $\gamma(\epsilon,t)$ in Subsection~\ref{subsec:gamma}.
Summing up the above contributions of all types of single-cycle neighborhood graphs, we obtain
\begin{multline*}
\xi(t)=\sum_{s_1=1}^{t-1}\sum_{s_2=2s_1+1}^{2t}\frac1{2}\lambda''(1)\rho'(1)^2(\lambda'(1)\rho'(1))^{s_2-s_1-2}\\
+\sum_{s_1=0}^{t-1}\sum_{s_2=2s_1+2}^{2t}\frac1{2}\rho''(1)\lambda'(1)(\lambda'(1)\rho'(1))^{s_2-s_1-2}\\
+\sum_{s=1}^{2t}\frac1{2}(\lambda'(1)\rho'(1))^s\\
=\frac1{2}\Biggl[
\lambda''(1)\rho'(1)^2\frac{(1-(\lambda'(1)\rho'(1))^{t-1})(1-(\lambda'(1)\rho'(1))^t)}{(1-\lambda'(1)\rho'(1))^2}\\
+\rho''(1)\lambda'(1)\frac{(1-(\lambda'(1)\rho'(1))^t)^2}{(1-\lambda'(1)\rho'(1))^2}\\
+\lambda'(1)\rho'(1)\frac{1-(\lambda'(1)\rho'(1))^{2t}}{1-\lambda'(1)\rho'(1)}\Biggr]\text{.}
\end{multline*}

It should be noted that the above result can alternatively be obtained 
via the generating function method described in the previous subsection.  
Indeed, since the probability of all neighborhood graphs is exactly $1$
and since the probability of neighborhood graphs which contain more than one cycle is $\Theta(n^{-2})$,
the coefficient of $n^{-1}$ in the probability of cycle-free neighborhood graphs is $-\xi(t)$, i.e.,
the probability of tree neighborhood graphs is $1-\xi(t)/n+\Theta(n^{-2})$.
Hence, the above result for the quantity $\xi(t)$ is obtained by enumeration of the coefficient of $n^{-1}$
in the probability of cycle-free neighborhood graphs $-\beta(1,t)$ using the generating function method in the previous subsection.

\begin{figure*}[!b]
\setcounter{equation}{11}
\hrulefill
\begin{align}
f(t,s,p) &:=
\begin{cases}
\epsilon, &\text{if } t = 0\\
\epsilon \frac{\lambda'(P_\epsilon(t))}{\lambda'(1)} g(t,s-1,p), &\text{otherwise}
\end{cases}\text{ ,\hspace{2em}}
g(t,s,p) :=
\begin{cases}
p, &\text{if } s = 0\\
1-\frac{\rho'(1-Q_\epsilon(t))}{\rho'(1)} (1-f(t-1,s,p)), &\text{otherwise}\\
\end{cases}\nonumber\\
G_1(t,s) &:= \begin{cases}
1, &\text{if } s = 0\\
\left(1-\frac{\rho'(1-Q_\epsilon(t))}{\rho'(1)}\right)^2+2\frac{\rho'(1-Q_\epsilon(t))}{\rho'(1)}\left(1-\frac{\rho'(1-Q_\epsilon(t))}{\rho'(1)}\right)f(t-1,s,1)\\
\hspace{24em}+\left(\frac{\rho'(1-Q_\epsilon(t))}{\rho'(1)}\right)^2G_2(t-1,s-1), &\text{otherwise}
\end{cases}\nonumber\\
G_2(t,s) &:= \begin{cases}
\epsilon\frac{\lambda'(P_\epsilon(t))}{\lambda'(1)}, &\text{if } s = 0\\
\left(\epsilon\frac{\lambda'(P_\epsilon(t))}{\lambda'(1)}\right)^2G_1(t,s-1), &\text{otherwise}
\end{cases}\nonumber\\
G_3(t,s) &:= \begin{cases}
1-\epsilon\frac{\lambda'(P_\epsilon(t))}{\lambda'(1)},&\text{if }s=0\\
1-2f(t,s+1,1)+ \left(\epsilon\frac{\lambda'(P_\epsilon(t))}{\lambda'(1)}\right)^2G_1(t,s-1),&\text{otherwise}
\end{cases}\nonumber\\
F_{v}(t,s_1,s_2) &= \frac1{2}\lambda''(1)\rho'(1)^2(\lambda'(1)\rho'(1))^{s_2-s_1-2}
Q_\epsilon(t+1)\nonumber\\
&\quad\times g\left(t,s_1-1,1-\frac{\rho'(1-Q_\epsilon(t-s_1+1))}{\rho'(1)}\left(1-\epsilon\frac{\lambda''(P_\epsilon(t-s_1))}{\lambda''(1)}G_1(t-s_1,s_2-2s_1-1)\right)\right)\label{f12}\\
F_{c}(t,s_1,s_2) &= \frac1{2}\rho''(1)\lambda'(1)(\lambda'(1)\rho'(1))^{s_2-s_1-2}
Q_\epsilon(t+1) g\left(t,s_1, 1-\frac{\rho''(1-Q_\epsilon(t-s_1))}{\rho''(1)}G_3(t-s_1-1,s_2-2s_1-2)\right)\label{f34}\\
F_{r}(t,s) &= \frac1{2}(\lambda'(1)\rho'(1))^{s}\epsilon\frac{\lambda'(P_\epsilon(t))}{\lambda'(1)}
G_1(t,s-1)\label{f56}
\end{align}
\setcounter{equation}{10}
\end{figure*}

\subsection{Single-cycle neighborhood graph ensembles}\label{subsec:scnge}
\textit{Single cycle neighborhood graph ensembles} are defined in this subsection 
in order to make the description of the calculation of $\gamma(\epsilon,t)$ in the next subsection more tractable.  
A single-cycle neighborhood graph ensemble for an arbitrary fixed type and parameters 
is defined not in terms of single-cycle neighborhood graphs
but in terms of representatives of their equivalence classes, 
with the specified type and parameters.  
The definition of single-cycle neighborhood graph ensembles is motivated by~\eqref{eq:enumcoef}.
The probability, to be defined in this subsection, of a representative, 
denoted as $G$ by a slight abuse of notation, 
can be considered as the large blocklength limit of the conditional probability, 
measured by the neighborhood graph ensemble, of the single-cycle neighborhood graphs in the equivalence class represented by $G$
conditioned on that a single-cycle neighborhood graph has a particular type and parameters.
The probability of a representative $G$ of an equivalence class of single-cycle neighborhood graphs in $\bar{\mathcal{S}}_v(t,s_1,s_2)$ is
\begin{multline}
\mathbb{P}_{v}^{(t,s_1,s_2)}(G)
:=
\frac{L_{|v_0|}|v_0|}{L'(1)}
\frac{\lambda_{|w|}(|w|-1)(|w|-2)}{\lambda''(1)}\\
\times\prod_{v\in \mathcal{V}(S(G))\backslash \{v_0, w\}} \frac{\lambda_{|v|}(|v|-1)}{\lambda'(1)}
\prod_{c\in \mathcal{C}(S(G))} \frac{\rho_{|c|}(|c|-1)}{\rho'(1)}\\
\times\prod_{v\in \mathcal{V}(G)\backslash \mathcal{V}(S(G))} \lambda_{|v|}
\prod_{c\in \mathcal{C}(G)\backslash\mathcal{C}(S(G))} \rho_{|c|}
\label{eq:pbsgl}
\end{multline}
where $w$ denotes the shallowest variable node in the cycle.
Similarly, the probability of a representative $G$ of an equivalence class of single-cycle neighborhood graphs in $\bar{\mathcal{S}}_c(t,s_1,s_2)$ is
\begin{multline*}
\mathbb{P}_{c}^{(t,s_1,s_2)}(G) :=
\frac{L_{|v_0|}|v_0|}{L'(1)}
\frac{\rho_{|s|}(|s|-1)(|s|-2)}{\rho''(1)}\\
\times\prod_{v\in \mathcal{V}(S(G))\backslash v_0} \frac{\lambda_{|v|}(|v|-1)}{\lambda'(1)}
\prod_{c\in \mathcal{C}(S(G))\backslash s} \frac{\rho_{|c|}(|c|-1)}{\rho'(1)}\\
\times\prod_{v\in \mathcal{V}(G)\backslash \mathcal{V}(S(G))} \lambda_{|v|}
\prod_{c\in \mathcal{C}(G)\backslash\mathcal{C}(S(G))} \rho_{|c|}
\end{multline*}
where $s$ denotes the shallowest check node in the cycle, and
the probability of $G\in\bar{\mathcal{S}}_r(t,s)$ is
\begin{multline*}
\mathbb{P}_{r}^{(t,s)}(G) :=
\frac{L_{|v_0|}|v_0|(|v_0|-1)}{L''(1)}\prod_{v\in \mathcal{V}(G)\backslash \mathcal{V}(S(G))} \lambda_{|v|}\\
\times \prod_{c\in \mathcal{C}(G)\backslash\mathcal{C}(S(G))} \rho_{|c|}
\prod_{v\in \mathcal{S}(G)\backslash v_0} \frac{\lambda_{|v|}(|v|-1)}{\lambda'(1)}
\prod_{c\in \mathcal{S}(G)} \frac{\rho_{|c|}(|c|-1)}{\rho'(1)}.
\end{multline*}
These ensembles are used in Subsection~\ref{subsec:gamma} for calculation of $\gamma(\epsilon,t)$.

\subsection{The contribution of single-cycle neighborhood graphs}\label{subsec:gamma}
The contribution $\gamma(\epsilon,t)$ of single-cycle neighborhood graphs can be decomposed according to the types and parameters of single-cycle
neighborhood graphs.
\begin{align*}
\gamma(\epsilon,t) &= 
\sum_{G\in\mathcal{S}_t} (\lim_{n\to\infty}n\Pn(G))\Pb(\epsilon,G)\\
&= \sum_{s_1=1}^{t-1}\sum_{s_2=2s_1+1}^{2t} \sum_{G\in\mathcal{S}_v(t,s_1,s_2)}(\lim_{n\to\infty}n\Pn(G))\Pb(\epsilon,G)\\
&\quad+\sum_{s_1=0}^{t-1} \sum_{s_2=2s_1+2}^{2t} \sum_{G\in\mathcal{S}_c(t,s_1,s_2)}(\lim_{n\to\infty}n\Pn(G))\Pb(\epsilon,G) \\ 
&\quad+\sum_{s=1}^{2t}\sum_{G\in\mathcal{S}_r(t,s)}(\lim_{n\to\infty}n\Pn(G))\Pb(\epsilon,G) \\
&=: \sum_{s_1=1}^{t-1}\sum_{s_2=2s_1+1}^{2t} F_{v}(t,s_1,s_2)\\
&\quad+\sum_{s_1=0}^{t-1} \sum_{s_2=2s_1+2}^{2t} F_{c}(t,s_1,s_2) + 
\sum_{s=1}^{2t} F_{r}(t,s)
\end{align*}
where $F_v(t,s_1,s_2)$, $F_c(t,s_1,s_2)$ and $F_r(t,s)$ are the contributions of
single-cycle neighborhood graphs in $\mathcal{S}_v(t,s_1,s_2)$, $\mathcal{S}_c(t,s_1,s_2)$ and $\mathcal{S}_r(t,s)$, respectively.
A set of formulas for calculations of these quantities are shown in the next theorem.

\begin{theorem}\label{gamma}
$\gamma(\epsilon,t)$ for $(\lambda(x),\rho(x))$-irregular ensembles is calculated as
\begin{multline*}
\gamma(\epsilon,t) = \sum_{s_1=1}^{t-1}\sum_{s_2=2s_1+1}^{2t} F_{v}(t,s_1,s_2)\\
+\sum_{s_1=0}^{t-1} \sum_{s_2=2s_1+2}^{2t} F_{c}(t,s_1,s_2) + 
\sum_{s=1}^{2t} F_{r}(t,s)
\end{multline*}
where $F_{v}(t,s_1,s_2)$, $F_{c}(t,s_1,s_2)$ and $F_{r}(t,s)$ are shown in (\ref{f12}), (\ref{f34}) and (\ref{f56}), respectively.
If $\lambda''(1)=0$, $F_v(t,s_1,s_2)$ is defined as $0$.
\end{theorem}
\setcounter{equation}{14}

A derivation of $F_v(t,s_1,s_2)$ is described in the following.
Similarly to~\eqref{eq:enumcoef},
the contribution $F_v(t,s_1,s_2)$ of neighborhood graphs in $\mathcal{S}_v(t,s_1,s_2)$ to $\gamma(\epsilon,t)$ is obtained as
\begin{align*}
&F_v(t,s_1,s_2)=\sum_{G\in\mathcal{S}_{v}(t,s_1,s_2)}\left(\lim_{n\to\infty}n\Pn(G)\right)\Pb(\epsilon,G)\\
&=\frac1{L'(1)}\sum_{G\in\bar{\mathcal{S}}_{v}(t,s_1,s_2)}
L_{|v_0|}|v_0|\nonumber\\
&\quad \times \prod_{v\in \mathcal{V}(S(G))\backslash \{v_0,w\}}\lambda_{|v|}(|v|-1) \prod_{c\in \mathcal{C}(S(G))} \rho_{|c|}(|c|-1)\nonumber \\
&\quad \times\lambda_{|w|}\binom{|w|-1}{2}\prod_{v\in \mathcal{V}(G)\backslash \mathcal{V}(S(G))}\lambda_{|v|} \prod_{c\in \mathcal{C}(G)\backslash \mathcal{C}(S(G))} \rho_{|c|}\nonumber \\
&\quad \times\Pb(\epsilon,G)\\ 
&=\frac1{2}\lambda''(1)\rho'(1)^2(\lambda'(1)\rho'(1))^{s_2-s_1-2}\\
&\quad \times\sum_{G\in\bar{\mathcal{S}}_{v}(t,s_1,s_2)}
\frac{L_{|v_0|} |v_0|}{L'(1)}\prod_{v\in \mathcal{V}(S(G))\backslash \{v_0,w\}}\frac{\lambda_{|v|}(|v|-1)}{\lambda'(1)}\\
&\quad\times \left(\prod_{c\in \mathcal{C}(S(G))} \frac{\rho_{|c|}(|c|-1)}{\rho'(1)}\right)
\frac{\lambda_{|w|}(|w|-1)(|w|-2)}{\lambda''(1)}\\
&\quad \times\prod_{v\in \mathcal{V}(G)\backslash \mathcal{V}(S(G))}\lambda_{|v|} \prod_{c\in \mathcal{C}(G)\backslash \mathcal{C}(S(G))} \rho_{|c|}
\Pb(\epsilon,G)\\
&=\frac1{2}\lambda''(1)\rho'(1)^2(\lambda'(1)\rho'(1))^{s_2-s_1-2}\\
&\quad \times\sum_{G\in\bar{\mathcal{S}}_{v}(t,s_1,s_2)}\mathbb{P}_{v}^{(t,s_1,s_2)}(G)\Pb(\epsilon,G)
\end{align*}
Hence, we have to calculate the expected error probability over a single-cycle neighborhood ensemble.
Marginalizing the non-cycle part of $S(G)$ and trees incident to them, if any, 
we obtain
\begin{align}
&F_v(t,s_1,s_2)=
\frac1{2}\lambda''(1)\rho'(1)^2(\lambda'(1)\rho'(1))^{s_2-s_1-2}\nonumber\\
&\times\sum_{Y,Z}
\prod_{v\in \mathcal{V}(Y)\backslash w}\frac{\lambda_{|v|}(|v|-1)}{\lambda'(1)} \prod_{c\in \mathcal{C}(Y)} \frac{\rho_{|c|}(|c|-1)}{\rho'(1)}\nonumber\\
&\quad \times\frac{\lambda_{|w|}(|w|-1)(|w|-2)}{\lambda''(1)}
\prod_{v\in \mathcal{V}(Z)}\lambda_{|v|} \prod_{c\in \mathcal{C}(Z)} \rho_{|c|}\nonumber\\
&\quad \times\epsilon\frac{L'(P_\epsilon(t))}{L'(1)}\bigg(1-\frac{\rho'(1-Q_\epsilon(t))}{\rho'(1)}\bigg(1-\epsilon\frac{\lambda'(P_\epsilon(t-1))}{\rho'(1)}\nonumber\\
&\quad \dotsm (1-p)\bigg)\bigg)
\label{gm1}
\end{align}
where $Y$ denotes the subgraph which consists of nodes in the cycle, where $Z$ denotes trees incident to $Y$, and
where $p$ denotes erasure probability of the message from $w$ to the shallow check node connected to $w$.
The calculation of a non-cycle part in~\eqref{gm1} is similar to the derivation of density evolution in Lemma~\ref{lem:debec}.
Equation (\ref{gm1}) is calculated as
\begin{align*}
&\frac1{2}\lambda''(1)\rho'(1)^2(\lambda'(1)\rho'(1))^{s_2-s_1-2}Q_\epsilon(t+1)\\
&\quad \times g\bigg(t,s_1-1,1-\frac{\rho'(1-Q_\epsilon(t-s_1+1))}{\rho'(1)}\bigg(1\\
&-\sum_{Y,Z}
\prod_{v\in \mathcal{V}(Y)\backslash w}\frac{\lambda_{|v|}(|v|-1)}{\lambda'(1)} \prod_{c\in \mathcal{C}(Y)} \frac{\rho_{|c|}(|c|-1)}{\rho'(1)}\\
&\quad \times\frac{\lambda_{|w|}(|w|-1)(|w|-2)}{\lambda''(1)}
\prod_{v\in \mathcal{V}(Z)}\lambda_{|v|} \prod_{c\in \mathcal{C}(Z)} \rho_{|c|} p\bigg)\bigg)
\text{.}
\end{align*}

Hence, if one can prove the equality 
\begin{multline}
\sum_{Y,Z}
 \prod_{v\in \mathcal{V}(Y)\backslash w}\frac{\lambda_{|v|}(|v|-1)}{\lambda'(1)} \prod_{c\in \mathcal{C}(Y)} \frac{\rho_{|c|}(|c|-1)}{\rho'(1)}\\
 \times\frac{\lambda_{|w|}(|w|-1)(|w|-2)}{\lambda''(1)}
 \prod_{v\in \mathcal{V}(Z)}\lambda_{|v|} \prod_{c\in \mathcal{C}(Z)} \rho_{|c|} p\\
 = \epsilon \frac{\lambda''(P_\epsilon(t-s_1))}{\lambda''(1)}G_1(t-s_1,s_2-2s_1-1)\label{ppp}
\end{multline}
then (\ref{f12}) will immediately be obtained.  

Now we prove~\eqref{ppp}.  
First, marginalizing $w$ and trees incident to $w$, denoted by $Z_w$, the left-hand side of (\ref{ppp}) is calculated as
\begin{multline}
\sum_{Y\backslash w, Z\backslash Z_w}
 \prod_{v\in \mathcal{V}(Y\backslash w)}\frac{\lambda_{|v|}(|v|-1)}{\lambda'(1)} \prod_{c\in \mathcal{C}(Y\backslash w)} \frac{\rho_{|c|}(|c|-1)}{\rho'(1)}\\
 \times\prod_{v\in \mathcal{V}(Z\backslash Z_w)}\lambda_{|v|} \prod_{c\in \mathcal{C}(Z\backslash Z_w)} \rho_{|c|} 
 \epsilon \frac{\lambda''(P_\epsilon(t-s_1))}{\lambda''(1)}q\label{cq}
\end{multline}
where $q$ denotes the probability that two messages into $w$ from the check nodes connected to $w$ in the cycle are both erased. 
Let $c_1$ and $c_2$ denote the check nodes in the cycle incident to $w$.
If $c_1$ and $c_2$ are the same, i.e., if $s_2-2s_1-1=0$ holds, then $q=1$.
Otherwise, $q$ is decomposed to four components as
\begin{multline}
q = P(e_1,e_2) = P(A_1,A_2,e_1,e_2) + P(\bar{A}_1,A_2,e_1,e_2)\\ + P(A_1,\bar{A}_2,e_1,e_2) + P(\bar{A}_1,\bar{A}_2,e_1,e_2)\\
= P(A_1)P(A_2) + P(e_1\mid \bar{A}_1, A_2)P(\bar{A}_1)P(A_2) \\
+ P(e_2\mid A_1, \bar{A}_2)P(A_1)P(\bar{A}_2) + P(\bar{A}_1,\bar{A}_2,e_1,e_2)\label{decq}
\end{multline}
where $e_1$ and $e_2$ denote the events that the messages from $c_1$ and $c_2$ to $w$ are erased, respectively,
and where $A_1$ and $A_2$ denote the events that at least one message from outside the cycle into $c_1$ and $c_2$
is erased, respectively.
Calculating the marginalization in (\ref{cq}), the first term in (\ref{decq}) becomes
\begin{equation}
\left(1-\frac{\rho'(1-Q_\epsilon(t-s_1))}{\rho'(1)}\right)^2\text{.}\label{q1}
\end{equation}
Each of the second and third terms becomes 
\begin{multline}
\frac{\rho'(1-Q_\epsilon(t-s_1))}{\rho'(1)}\left(1-\frac{\rho'(1-Q_\epsilon(t-s_1))}{\rho'(1)}\right)\\
\times f(t-s_1-1,s_2-2s_1-1,1)\text{.}\label{q2}
\end{multline}
At last, the fourth term becomes
\begin{equation}
\left(\frac{\rho'(1-Q_\epsilon(t-s_1))}{\rho'(1)}\right)^2 r\label{q3}
\end{equation}
where $r$ denotes the probability that both of messages to $c_1$ and $c_2$ from variable nodes $v_1$ and $v_2$ 
in the cycle which connect to $c_1$ and $c_2$, respectively, and which are not $w$, are erased.
If $v_1$ and $v_2$ are the same, i.e., if $s_2-2s_2-1=1$ holds, 
then $r=\epsilon\lambda'(P_\epsilon(t-s_1-1))/\lambda'(1)$.
Otherwise,
\begin{equation*}
r=\left(\epsilon\frac{\lambda'(P_\epsilon(t-s_1-1))}{\lambda'(1)}\right)^2q'
\end{equation*}
where $q'$ denotes the probability that both of messages to $v_1$ and $v_2$
from check nodes $c_3$ and $c_4$
in the cycle which connect to $v_1$ and $v_2$, respectively, and which are not $c_1$ and $c_2$, respectively,
are erased.
The probability $q'$ is obtained in the same way as $q$.
Summing (\ref{q1}), (\ref{q2}) and (\ref{q3}), we obtain $G_1(t-s_1,s_2-2s_1-1)$.
Hence, we obtain (\ref{ppp}) and the proof that the contribution of neighborhood graphs
of Type I and Type II with the parameters $s_1$ and $s_2$ is $F_v(t,s_1,s_2)$ is done.



In almost the same way, the contributions of neighborhood graphs of Type III and Type IV
are obtained as
\begin{multline*}
\frac1{2}\rho''(1)\lambda'(1)(\lambda'(1)\rho'(1))^{s_2-s_1-2} Q_\epsilon(t+1)\\
\times g\bigg(t,s_1, 1-\frac{\rho''(1-Q_\epsilon(t-s_1))}{\rho''(1)}\\
\times G_3(t-s_1-1,s_2-2s_1-2)\bigg)
\end{multline*}
and that of Type V and Type VI are obtained as
\begin{equation*}
\frac1{2}(\lambda'(1)\rho'(1))^{s}\epsilon\frac{\lambda'(P_\epsilon(t))}{\lambda'(1)} G_1(t,s-1)\text{.}
\end{equation*}
Since the derivation is similar, the proof is omitted.

\section{The limit of $\alpha(\epsilon,t)$}\label{sec:limalpha}
In this section, the limit values $\alpha(\epsilon,\infty) := \lim_{t\to\infty} \alpha(\epsilon,t)$ for regular ensembles are shown.
The limit $\alpha(\epsilon,\infty)$ has a simple expression while the expression of $\alpha(\epsilon,t)$ is complicated and recursive.
Empirically, the approximation using $\alpha(\epsilon,\infty)$ instead of $\alpha(\epsilon,t)$ is accurate even for small blocklength
if $\epsilon$ is close to $0$ or $1$, as will be observed in Section~\ref{sec:cal}.
The proof of the following theorem is in Appendix~\ref{anbp}.

\begin{theorem}\label{limbp}
For $(l,r)$-regular ensembles, 
let
\begin{align*}
P_\epsilon(\infty)&:=\lim_{t\to\infty}P_\epsilon(t)\\
Q_\epsilon(\infty)&:=\lim_{t\to\infty}Q_\epsilon(t)\\
p&:= \epsilon (l-1) P_\epsilon(\infty)^{l-2}\\
q&:= (r-1) (1-Q_\epsilon(\infty))^{r-2}\\
v&:= \epsilon (l-1)(l-2) P_\epsilon(\infty)^{l-3}\\
w &:= (r-1)(r-2) (1-Q_\epsilon(\infty))^{r-3}\text{.}
\end{align*}
If $pq<1$, the limit is
\begin{multline*}
\alpha(\epsilon,\infty) =
\frac1{2}\frac1{1-pq}\left(pq+Q_\epsilon(\infty)\frac1{1-pq}q^2v\right)\\
\times\left[\frac1{1-pq}(P_\epsilon(\infty)-Q_\epsilon(\infty)) + 1-P_\epsilon(\infty) Q_\epsilon(\infty)\right]\\
+\frac1{2}Q_\epsilon(\infty)\frac1{(1-pq)^2}wp\\
\times\left[\frac1{1-pq}(Q_\epsilon(\infty)-P_\epsilon(\infty))+(1-P_\epsilon(\infty))(1- Q_\epsilon(\infty))\right].
\end{multline*}
\end{theorem}
The quantity $pq$ which appears in the condition of the theorem is the slope of the function of density evolution $f_\text{de}(x) = \epsilon\lambda(1-\rho(1-x))$,
which described the evolution of $Q_\epsilon(t)$ in Lemma~\ref{lem:debec},
at the largest fixed point $x=Q_\epsilon(\infty)\in[0,1]$, where $\lambda(x) := x^{l-1}$ and $\rho(x) := x^{r-1}$.
Hence, $pq\le1$ is always satisfied. $pq=1$ if and only if $y=f_\text{de}(x)$ touches $y=x$ at the largest fixed point.
Such points of $\epsilon$ includes the threshold $\epsilon_\text{BP}$ and the discontinuous point of the largest fixed point with respect to $\epsilon$.

Especially, below the threshold, 
\begin{equation*}
\alpha(\epsilon,\infty) = \frac1{2}\frac{\epsilon\lambda'(0)\rho'(1)}{1-\epsilon\lambda'(0)\rho'(1)}.
\end{equation*}
This quantity also appears in \eqref{eq:appinf}.
This fact implies that the following two limits are equal below the threshold for regular ensembles.
\begin{multline*}
\lim_{t\to\infty}\lim_{n\to\infty} n(\Pb(n,\epsilon,t)-\Pb(\infty,\epsilon,t))\\
=\lim_{n\to\infty}\lim_{t\to\infty} n(\Pb(n,\epsilon,t)-\Pb(\infty,\epsilon,t))
\end{multline*}

The limit $\alpha(\epsilon,\infty)$ for irregular ensembles is an open problem.

\section{Uniform convergence under fixed number of iterations}\label{sec:uniconv}
As mentioned in the introduction, 
the bit error probability after infinite number of iterations converges to a discontinuous curve
with respect to erasure probability of a channel as the blocklength tends to infinity if $\lambda'(0)\rho'(1)\epsilon_{\text{BP}}<1$.
Since the bit error probability for finite blocklength is continuous, the convergence is not uniform.
Due to the lack of uniform convergence, an approximation \eqref{eq:appinf} using asymptotic expansion with respect to blocklength
is not accurate near the discontinuous points. 
Hence, for accurate approximation near discontinuous points,
other approximations should be considered.
The scaling-law-based approximation method was introduced by Amraoui et al.~\cite{4777618}, \cite{Am07} 
for this purpose.

In this section, we will show that
the bit error probability after a fixed number of iterations converges to a limit uniformly
in contrast to the case of infinite number of iterations,
which immediately implies that the approximation \eqref{eq:apx}
is accurate for all $\epsilon$ uniformly when the blocklength is sufficiently large.
We have to show
\begin{equation}
\left|\Pb(n,\epsilon,t) - \Pb(\infty,\epsilon,t)\right| \le C
\label{unf}
\end{equation}
where $C=o(1)$ as $n\to\infty$ and $C$ does not depend on $\epsilon$.
The left-hand side of (\ref{unf}) is bounded as
\begin{multline}
\left|\Pb(n,\epsilon,t) - \Pb(\infty,\epsilon,t)\right| = \bigg|\sum_{G\in\mathcal{T}_t}\Pn(G)\Pb(\epsilon,G)\\
+\sum_{G\in\mathcal{S}_t}\Pn(G)\Pb(\epsilon,G)
+\sum_{G\in\mathcal{G}_t\backslash(\mathcal{T}_t\cup\mathcal{S}_t)}\Pn(G)\Pb(\epsilon,G)\\
-\Pb(\infty,\epsilon,t)\bigg|\\
\le\left|
\sum_{G\in\mathcal{T}_t}\Pn(G)\Pb(\epsilon,G)-\Pb(\infty,\epsilon,t)-\beta(\epsilon,t)\frac1{n}\right|\\
+\left|\sum_{G\in\mathcal{S}_t}\Pn(G)\Pb(\epsilon,G)-\gamma(\epsilon,t)\frac1{n}\right|\\
+\left|\alpha(\epsilon,t)\right|\frac1{n}+\sum_{G\in\mathcal{G}_t\backslash(\mathcal{T}_t\cup\mathcal{S}_t)}\Pn(G)\text{.}\label{eq:unf}
\end{multline}
From Lemma \ref{kcycles}, the last term in the rightmost side of~(\ref{eq:unf}), 
which depends on $t$ but not on $\epsilon$, is $\Theta(n^{-2})$.
The first term in the rightmost side of~(\ref{eq:unf}) is bounded as
\begin{multline*}
\left|\sum_{G\in\mathcal{T}_t}\Pn(G)\Pb(\epsilon,G)-\Pb(\infty,\epsilon,t)-\beta(\epsilon,t)\frac1{n}\right|\\
=\Biggl|\sum_{G\in\mathcal{T}_t}\left(\Pn(G)-\mathbb{P}_\infty(G)-\frac1n\left[\lim_{n\to\infty} n\left(\Pn(G)-\mathbb{P}_\infty(G)\right)\right]\right)\\
\times\Pb(\epsilon,G)\Biggr|\\
\le\sum_{G\in\mathcal{T}_t}\left|\left(\Pn(G)-\mathbb{P}_\infty(G)
-\frac1n\left[\lim_{n\to\infty} n\left(\Pn(G)-\mathbb{P}_\infty(G)\right)\right]\right)\right|
\end{multline*}
Similarly, the second term in the rightmost side of~(\ref{eq:unf}) is also bounded as
\begin{multline*}
\left|\sum_{G\in\mathcal{S}_t}\Pn(G)\Pb(\epsilon,G)-\gamma(\epsilon,t)\frac1{n}\right|\\
=\left|\sum_{G\in\mathcal{S}_t}\left(\Pn(G)-\frac1n\left[\lim_{n\to\infty}n\Pn(G)\right]\right)\Pb(\epsilon,G)\right|\\
\le\sum_{G\in\mathcal{S}_t}\left|\Pn(G)-\frac1n\left[\lim_{n\to\infty}n\Pn(G)\right]\right|
\end{multline*}
The above two bounds are $\Theta(n^{-2})$ and are independent of $\epsilon$.
Hence, (\ref{eq:unf}) is upper bounded by
\begin{equation*}
\left|\alpha(\epsilon,t)\right|\frac1{n} + D
\end{equation*}
where $D=\Theta(n^{-2})$ depends on $t$ but not on $\epsilon$.

Since $|\alpha(\epsilon,t)|$ is continuous on $\epsilon\in [0,1]$ and so bounded,
we conclude that the bit error probability under a finite number of iterations converges to the limit uniformly
as the blocklength tends to infinity.
More accurately, we obtain
\begin{equation}
\left|\Pb(n,\epsilon,t) - \Pb(\infty,\epsilon,t) - \alpha(\epsilon,t)\frac1{n}\right| \le D\label{dif}
\end{equation}
from the above results.
Equation (\ref{dif}) bounds an error of the approximation.
However, this bound is available only under the assumption that the blocklength is sufficiently large so that all possible neighborhood graphs could be generated.
In the next section, we observe via numerical calculations and simulations that the approximation is also accurate even if one cannot expect the assumption to be satisfied.

\begin{figure}[p]
\psfrag{e}{$\epsilon$}
\psfrag{alpha}{$\alpha(\epsilon,t)$}
\includegraphics[width=\hsize]{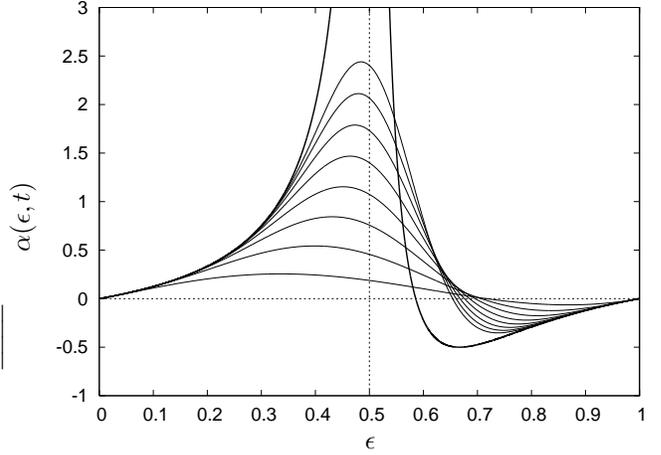}
\caption{Calculation results for $(2,3)$-regular ensemble. Thin curves show $\alpha(\epsilon,t)$ for $t=1,2,\dotsc,8$. Thick curve shows the limit $\alpha(\epsilon,\infty)$. The threshold $\epsilon_{\rm BP}$ is $0.5$.}
\label{23alpha}
\end{figure}

\begin{figure}[p]
\psfrag{e}{$\epsilon$}
\psfrag{alpha}{$\alpha(\epsilon,t)$}
\includegraphics[width=\hsize]{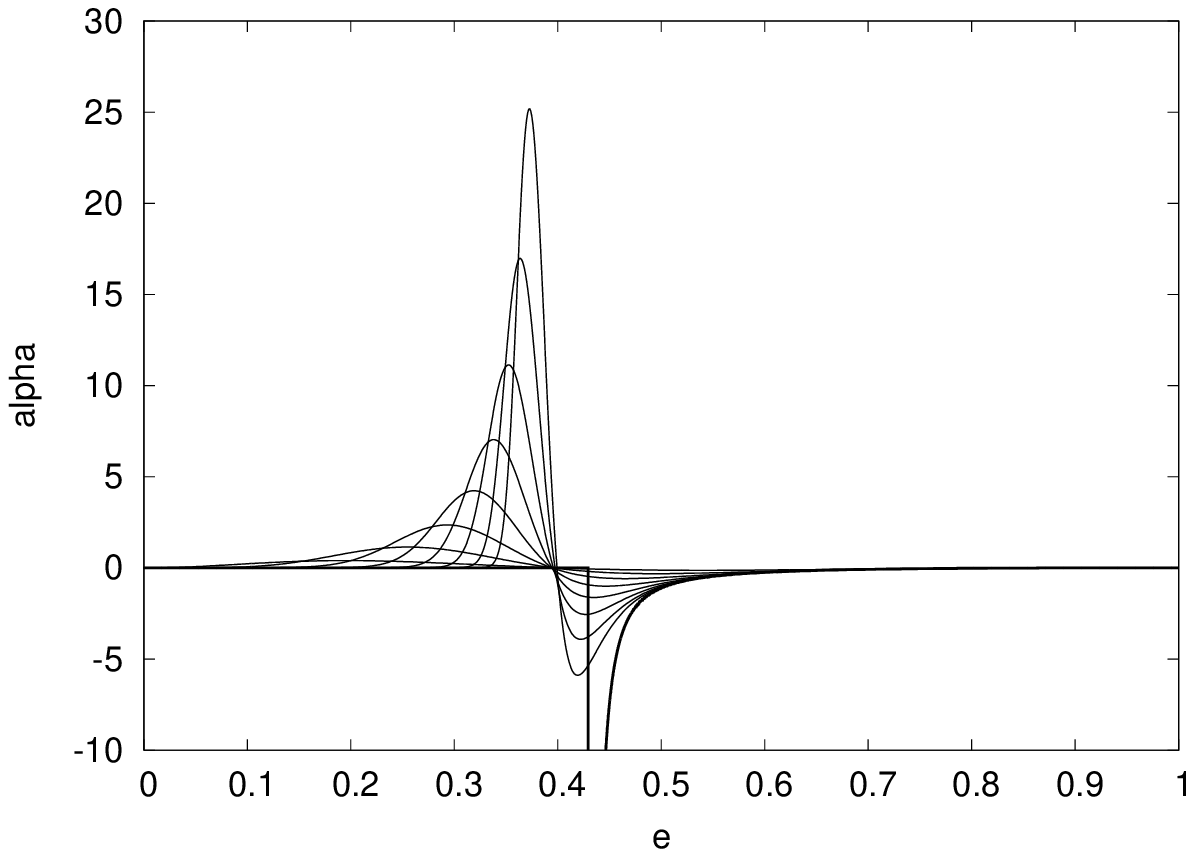}
\caption{Calculation results for $(3,6)$-regular ensemble. Thin curves show $\alpha(\epsilon,t)$ for $t=1,2,\dotsc,8$. Thick curve shows the limit $\alpha(\epsilon,\infty)$. The threshold $\epsilon_\mathrm{BP}$ is about $0.42944$.}
\label{36alpha}
\end{figure}

\begin{figure}[p]
\psfrag{e}{$\epsilon$}
\psfrag{alpha}{$\alpha(\epsilon,t)$}
\includegraphics[width=\hsize]{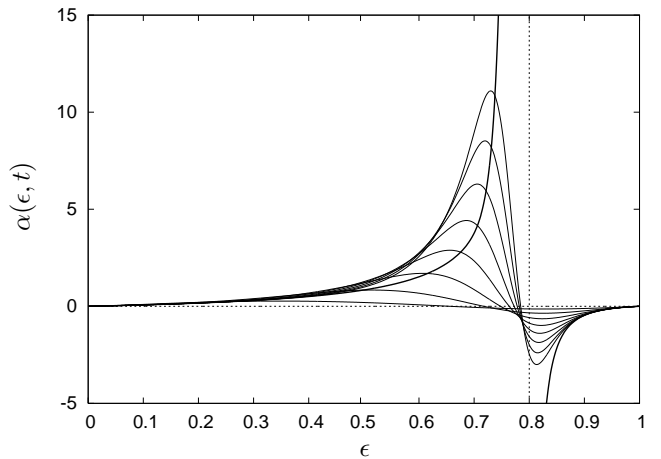}
\caption{Calculation results for an irregular ensemble.
$\lambda(x) = 0.500 x + 0.153 x^2 + 0.112 x^3 + 0.055 x^4 + 0.180 x^8$,
$\rho(x) = 0.492 x^2 + 0.508 x^3$.
Thin curves show $\alpha(\epsilon,t)$ for $t=1,2,\dotsc,8$. Thick curve shows the result with $t=50$. 
The threshold $\epsilon_\mathrm{BP}$ is about $0.8$.
}
\label{irgalpha}
\end{figure}

\begin{figure}[p]
\psfrag{epsilon}{$\epsilon$}
\psfrag{alpha}{$n|\Pb(n,\epsilon,t)-\Pb(\infty,\epsilon,t)|$}
\psfrag{al}{\footnotesize \hspace{-2em}$\alpha(\epsilon,t)$}
\includegraphics[width=\hsize]{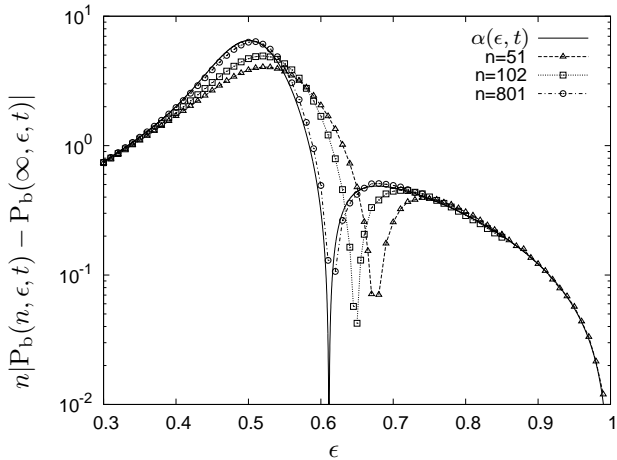}
\caption{Simulation results for $(2,3)$-regular ensemble.
Blocklengths are  $51$, $102$ and $801$.
Number of iterations is $20$.
}
\label{23sim}
\end{figure}

\begin{figure}[p]
\psfrag{epsilon}{$\epsilon$}
\psfrag{alpha}{$n|\Pb(n,\epsilon,t)-\Pb(\infty,\epsilon,t)|$}
\psfrag{al}{\footnotesize \hspace{-2em}$\alpha(\epsilon,t)$}
\includegraphics[width=\hsize]{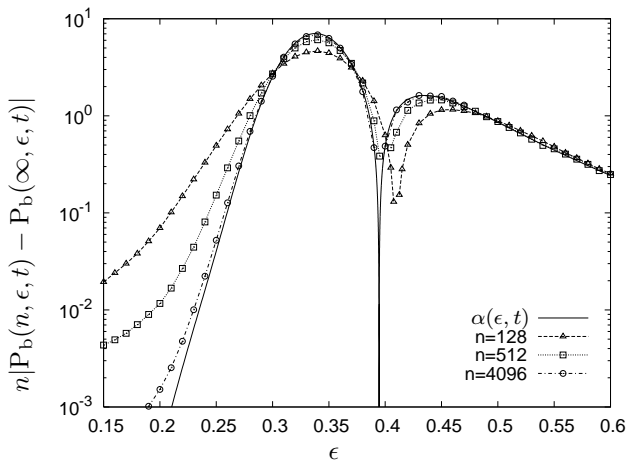}
\caption{Simulation results for $(3,6)$-regular ensemble.
Blocklengths are  $128$, $512$ and $4096$.
Number of iterations is $5$.
}
\label{36sim}
\end{figure}

\begin{figure}[p]
\psfrag{e}{$\epsilon$}
\psfrag{y}{$n|\Pb(n,\epsilon,t)-\Pb(\infty,\epsilon,t)|$}
\psfrag{alpha}{\footnotesize \hspace{-2em}$\alpha(\epsilon,t)$}
\includegraphics[width=\hsize]{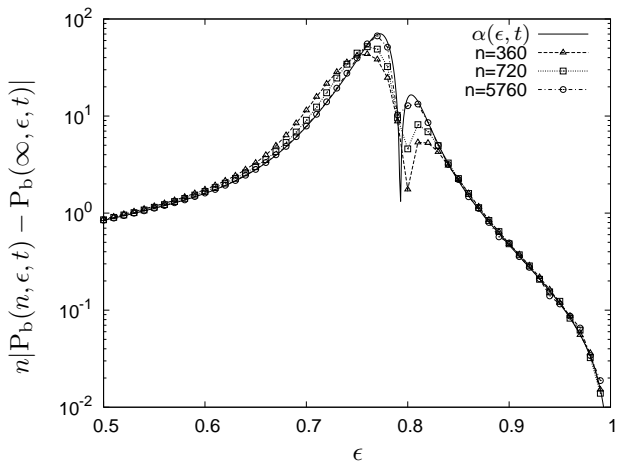}
\caption{Simulation results for an irregular ensemble.
$\lambda(x) = 0.500 x + 0.153 x^2 + 0.112 x^3 + 0.055 x^4 + 0.180 x^8$,
$\rho(x) = 0.492 x^2 + 0.508 x^3$.
Blocklengths are  $360$, $720$ and $5760$.
Number of iterations is $20$.
}
\label{irgsim}
\end{figure}

\section{Numerical calculations and simulations}\label{sec:cal}
In this section, we show calculation results of $\alpha(\epsilon,t)$ and $\alpha(\epsilon,\infty)$
and show simulation results of $n|\Pb(n,\epsilon,t) - \Pb(\infty,\epsilon,t)|$, 
the quantity which tends to $|\alpha(\epsilon,t)|$ as $n$ tends to infinity.

The results of calculating $\alpha(\epsilon,t)$ for the $(2,3)$-regular ensemble, the $(3,6)$-regular ensemble and an irregular ensemble are shown in Fig.~\ref{23alpha}, Fig.~\ref{36alpha} and Fig.~\ref{irgalpha}, respectively.
The coefficient $\alpha(\epsilon,t)$ seems to approach the limit $\alpha(\epsilon,\infty)$ quickly 
where $\epsilon$ is close to $0$ or $1$.

If $\epsilon$ satisfies the two conditions $\lambda'(0)\rho'(1)\epsilon < (\lambda'(1)\rho'(1))^{-1}$ and $\epsilon < \epsilon_\mathrm{BP}$,
then $\beta(\epsilon,t)$ tends to zero and $\gamma(\epsilon,t)$ tends to
the limit $\alpha(\epsilon,\infty)$ as $t$ tends to infinity.
In this case, we can understand intuitively that dominant events of decoding error are 
events of errors of channel outputs in single-cycle neighborhood graphs consisting of variable nodes of degree $2$ and check nodes,
i.e., single-cycle stopping sets equivalent to single-cycle codewords also discussed in~\cite{Mon06}.
However, if $\lambda'(0)\rho'(1)\epsilon > (\lambda'(1)\rho'(1))^{-1}$, 
which is the case when $\epsilon>0.25$ in Fig.~\ref{23alpha} and when $\epsilon>0.113\,48$ in Fig.~\ref{irgalpha}, 
even below the threshold,
$\beta(\epsilon,t)$ grows to $-\infty$ and $\gamma(\epsilon,t)$ grows to $+\infty$
exponentially in $t$.
The reason of this large cancellation between $\beta(\epsilon,t)$ and $\gamma(\epsilon,t)$ is not sufficiently understood.

Because of the large cancellation, multiprecision arithmetic was necessary in our calculations 
to avoid cancellation errors in computation of $\alpha(\epsilon,t)$ with large $t$.

Simulation results for the above ensembles are shown in Fig.~\ref{23sim}, Fig.~\ref{36sim} and Fig.~\ref{irgsim}, respectively.
For the $(2,3)$-regular ensemble, 
the simulation results with $n=801$ almost converge to $\alpha(\epsilon,t)$ for all $\epsilon$, 
as shown in Fig.~\ref{23sim}.  
It is also the case with the irregular ensemble which has variable nodes of degree $2$ (Fig.~\ref{irgsim}), 
where the simulation results with $n=5760$ are observed to converge well to $\alpha(\epsilon,t)$ for all $\epsilon$.  
For the $(3,6)$-regular ensemble (Fig.~\ref{36sim}), the simulation results almost converge to $\alpha(\epsilon,t)$ for $\epsilon>0.25$ at $n=4096$.
The agreement between simulation results and theoretical results is strange, 
since the pairs of the blocklength and the number of iterations are not suitable for density evolution technique
in which one assumes that neighborhood graphs are tree with high probability.
Indeed, the numbers of variable nodes in tree graphs are well above the total numbers of variable nodes 
in all the three cases considered: 
The number of variable nodes in a tree graph of depth $20$ in the $(2,3)$-regular ensemble is $4\,194\,302$ which is much greater than the blocklength $801$.
The number of variable nodes in a tree graph of depth $5$ in the $(3,6)$-regular ensemble is $166\,666$ which is much greater than the blocklength $4096$.
The number of variable nodes in the minimum tree graph of depth $20$ in the irregular ensemble is $4\,194\,302$ which is much greater than the blocklength $5760$.
We have not succeeded in finding an appropriate explanation to the observed quick convergence.

For $(3,6)$-regular ensemble, the convergence to $\alpha(\epsilon,t)$ is not fast for $\epsilon<0.25$.
In the low-$\epsilon$ region, dominant error events after infinite number of iterations are those induced by small stopping sets.
$(3,6)$-regular ensemble does not contain single-cycle stopping sets but 
contains three double-cycle stopping sets.
When $\epsilon$ is close to $0$, 
unless the blocklength is sufficiently large,
the bit error probability after a small number of iterations
is almost the same as that after infinite number of iterations,
since decoding will succeed after a few number of iterations with high probability.
It is also the case when $\epsilon$ is close to $1$, 
in which case decoding will fail after a few number of iterations with high probability.
Hence, in the low-$\epsilon$ region, the bit error probability decays like $\Theta(n^{-2})$ rather than $\Theta(n^{-1})$
unless the blocklength is sufficiently large.

The well-established fact that the bit error probability at error floor is well approximated by \eqref{eq:appinf}~\cite{Mon06} is interpreted as the statement that 
the bit error probability $\Pb(n,\epsilon,t)$ when $\epsilon$ is close to 0 and $\lambda'(0)>0$
is well approximated by $\Pb(\infty,\epsilon,t)+\alpha(\epsilon,\infty)/n$ for large $n$.  
From the observed quick convergence of $\alpha(\epsilon ,t)$ to $\alpha(\epsilon, \infty)$ and
that of $n(\Pb(n,\epsilon,t)-\Pb(\infty,\epsilon,t))$ to $\alpha(\epsilon ,t)$ for $\epsilon$ close to 1,
the same statement is empirically valid when $\epsilon$ is close to 1 as well.


\section{Conclusion}\label{sec:conc}
We have obtained the coefficient $\alpha(\epsilon,t)$ of the second dominant term 
in the asymptotic expansion of the bit error probability after a fixed number of iterations for irregular ensembles.
Furthermore, we have obtained the limit $\alpha(\epsilon,\infty)$ for regular ensembles.
At last, we have confirmed that approximations using $\alpha(\epsilon,t)$ are accurate even for small blocklength.

There are two important open problems.
The first one is the large cancellation problem between $\beta(\epsilon,t)$ and $\gamma(\epsilon,t)$.
The underlying mechanism of this cancellation has not been understood sufficiently, 
so that, for example, we do not know whether similar cancellations occur in higher-order terms.
The second one is the fast convergence problem of $\alpha(\epsilon,t)$.
Simulation results show that the convergence to $\alpha(\epsilon,t)$ is very fast.
This fact is strange since neighborhood graphs should include many cycles in moderate blocklengths.

Some other works remain to be done. 
First, the limit $\alpha(\epsilon,\infty)$ for irregular ensemble has not been derived.
Second, optimization of finite-length irregular and expurgated ensembles
given the number of iterations, blocklength, erasure probability, and allowable error probability,
similar to the finite-blocklength optimization by Amraoui et al.~\cite{4777618,Am07} 
for infinite number of iterations, 
is practically important.
Third, derivation of the coefficients of higher-order terms $n^{-2},n^{-3},\dotsc$ is an interesting problem.
Fourth, other limits may also be important in practice.  
An example is the limit of blocklength and the number of iterations tending to infinity simultaneously.  
Assume $t=c\log n / \log(\lambda'(1)\rho'(1))$ for some constant $c > 0$.
Then the probability of cycle-free neighborhood graphs tends to $1$ for $c<1/2$ and tends to $0$ for $c>1/2$~\cite{mezard2009ipa}.
It means that the cycle-free assumption is applicable only for $c<1/2$, 
so that methods like density evolution under $c>1/2$ are not available.
At last, generalization of the methods to general channels and BP or other message passing decoders is important.
There is a technically difficult problem due to reuse of messages from the same edges for calculation of the contributions
of single-cycle neighborhood graphs.

\appendices

\section{Proof of Theorem~\ref{limbp}}\label{anbp}
First, we show an alternative expression of $\alpha(\epsilon,t)$ for regular ensembles.
The new expression is useful for proving Theorem~\ref{limbp}.
Furthermore, the expression does not require multiprecision arithmetic which the previous expression requires in order to avoid the cancellation errors
in the calculation of the sum $\beta(\epsilon,t) + \gamma(\epsilon,t)$.
\begin{lemma}\label{lem:alpred}
For $(l,r)$-regular ensemble, $\alpha(\epsilon,t)$ is calculated as
\begin{multline*}
\alpha(\epsilon,t) = \sum_{s_1=1}^{t-1}\sum_{s_2=2s_1+1}^{2t} T_{v}(t,s_1,s_2)\\
+\sum_{s_1=0}^{t-1} \sum_{s_2=2s_1+2}^{2t} T_{c}(t,s_1,s_2)
+\sum_{s=1}^{2t} T_{r}(t,s)
\end{multline*}
where
\begin{multline*}
T_v(t,s_1,s_2) := \frac1{2}Q_\epsilon(t+1)\rho'(1-Q_\epsilon(t))\epsilon\lambda''(P_\epsilon(t-s_1))\\
\times \left(\prod_{k=1}^{s_1-1} \epsilon\lambda'(P_\epsilon(t-k))\rho'(1-Q_\epsilon(t-k))\right)\\
\times H_1(t-s_1,s_2-2s_1-1)
\end{multline*}
\begin{multline*}
T_c(t,s_1,s_2) := \frac1{2}Q_\epsilon(t+1)\rho''(1-Q_\epsilon(t-s_1))\\
\times \left(\prod_{k=0}^{s_1-1} \epsilon\lambda'(P_\epsilon(t-k-1))\rho'(1-Q_\epsilon(t-k))\right)\\
\times H_3(t-s_1-1,s_2-2s_1-2)
\end{multline*}
\begin{equation*}
T_r(t,s):=\frac1{2}\epsilon\lambda'(P_\epsilon(t))H_1(t,s-1)
\end{equation*}
\begin{multline*}
H_1(t,s) :=\\
\begin{cases}
\rho'(1)(1-P_\epsilon(t)^2),&\text{if } s = 0\\
\rho'(1-Q_\epsilon(t))^2 H_2(t-1,s-1),& \text{if } s\ge t\\
2(\rho'(1)-\rho'(1-Q_\epsilon(t)))(1-P_\epsilon(t-s))\\
\;\times\prod_{k=0}^{s-1} \rho'(1-Q_\epsilon(t-k)) \epsilon \lambda'(P_\epsilon(t-k-1))\\
\;+ \rho'(1-Q_\epsilon(t))^2 H_2(t-1,s-1),&\text{otherwise}
\end{cases}
\end{multline*}
\begin{equation*}
H_2(t,s) := \begin{cases}
\epsilon \lambda'(P_\epsilon(t))-\lambda'(1)Q_\epsilon(t+1)^2,&\text{if } s = 0\\
(\epsilon \lambda'(P_\epsilon(t)))^2 H_1(t,s-1),
\end{cases}
\end{equation*}
\begin{multline*}
H_3(t,s):=\\
\begin{cases}
\epsilon \lambda'(P_\epsilon(t))-\lambda'(1)Q_\epsilon(t+1)(2-Q_\epsilon(t+1)) ,&\text{if } s = 0\\
-(\epsilon \lambda'(P_\epsilon(t)))^2 H_1(t,s-1),&\text{if } s\ge t\\
2\epsilon\lambda'(P_\epsilon(t))(1-P_\epsilon(t-s))\\
\;\times\prod_{k=0}^{s-1} \rho'(1-Q_\epsilon(t-k)) \epsilon \lambda'(P_\epsilon(t-k-1))\\
\;- (\epsilon \lambda'(P_\epsilon(t)))^2 H_1(t,s-1),& \text{otherwise.}\\
\end{cases}
\end{multline*}
and where $\lambda(x) = x^{l-1}$ and $\rho(x) = x^{r-1}$.
\end{lemma}

\begin{IEEEproof}[Outline of proof of Lemma~\ref{lem:alpred}]
For $(l,r)$-regular ensemble, the cycle-free neighborhood graph is unique.
The coefficient of $n^{-1}$ in the probability of the unique cycle-free neighborhood graph is
\begin{equation*}
-\frac1{2}l(r-1)\frac{1-\{(l-1)(r-1)\}^t}{1-(l-1)(r-1)}\{(l-1)(r-1)\}^t\text{.}
\end{equation*}
Hence, $\beta(\epsilon,t)$ for $(l,r)$-regular ensemble is obtained as
\begin{multline*}
\beta(\epsilon,t) =
-\frac1{2}l(r-1)\frac{1-\{(l-1)(r-1)\}^t}{1-(l-1)(r-1)}\{(l-1)(r-1)\}^t\\ \times\epsilon P_\epsilon(t)^l\text{.}
\end{multline*}
It is decomposed as follows.
\begin{multline*}
\beta(\epsilon,t) =
-\epsilon P_\epsilon(t)^l\\
\times \frac1{2}\Bigg[\sum_{s_1=1}^{t-1}\sum_{2s_1+1}^{2t} \lambda''(1)\rho'(1)(\lambda'(1)\rho'(1))^{s_2-s_1-2}\\
+\sum_{s_1=0}^{t-1}\sum_{s_2=2s_1+2}^{2t}\rho''(1)\lambda'(1)(\lambda'(1)\rho'(1))^{s_2-s_1-2}\\
+ \sum_{s=1}^{2t} (\lambda'(1)\rho'(1))^s\Bigg].
\end{multline*}
Hence, $\alpha(\epsilon,t)$ is calculated as
\begin{align*}
&\alpha(\epsilon,t) = \sum_{s_1=1}^{t-1}\sum_{2s_1+1}^{2t}\\
& \left(F_v(t,s_1,s_2) - \frac1{2}\lambda''(1)\rho'(1)(\lambda'(1)\rho'(1))^{s_2-s_1-2}\epsilon P_\epsilon(t)^l\right)\\
&\quad +\sum_{s_1=0}^{t-1}\sum_{s_2=2s_1+2}^{2t}\\
&\left(F_c(t,s_1,s_2)-\frac1{2}\rho''(1)\lambda'(1)(\lambda'(1)\rho'(1))^{s_2-s_1-2}\epsilon P_\epsilon(t)^l\right)\\
&\quad + \sum_{s=1}^{2t}\left(F_r(t,s)- \frac1{2}(\lambda'(1)\rho'(1))^s\epsilon P_\epsilon(t)^l\right)\\
&=
\sum_{s_1=1}^{t-1}\sum_{s_2=2s_1+1}^{2t} T_{v}(t,s_1,s_2)
+\sum_{s_1=0}^{t-1} \sum_{s_2=2s_1+2}^{2t} T_{c}(t,s_1,s_2)\\
&\quad +\sum_{s=1}^{2t} T_{r}(t,s)\text{.}
\end{align*}
We omit calculations of $T_v(t,s_1,s_2)$, $T_c(t,s_1,s_2)$ and $T_r(t,s)$.
\end{IEEEproof}

\begin{IEEEproof}[Proof of Theorem~\ref{limbp}]
After some calculations, we obtain
\begin{multline*}
\lim_{t\to\infty}\sum_{s_1=1}^{t-1}\sum_{s_2=2s_1+1}^{2t} \lim_{u\to\infty} T_{v}(u,s_1,s_2) \\
= \frac1{2}Q_\epsilon(\infty)\frac1{(1-pq)^2}q^2v\\
\times\left[\frac1{1-pq}(P_\epsilon(\infty)-Q_\epsilon(\infty))+1-P_\epsilon(\infty)Q_\epsilon(\infty)\right]
\end{multline*}
\begin{multline*}
\lim_{t\to\infty}\sum_{s_1=0}^{t-1}\sum_{s_2=2s_1+2}^{2t} \lim_{u\to\infty} T_{c}(u,s_1,s_2) \\
= \frac1{2}Q_\epsilon(\infty)\frac1{(1-pq)^2}wp\\
\times\left[\frac1{1-pq}(Q_\epsilon(\infty)-P_\epsilon(\infty))+(1-P_\epsilon(\infty))(1-Q_\epsilon(\infty))\right]
\end{multline*}
\begin{multline*}
\lim_{t\to\infty}\sum_{s=1}^{2t} \lim_{u\to\infty} T_r(u,s) = \frac1{2}\frac1{1-pq}pq\\
\times\left[\frac1{1-pq}(P_\epsilon(\infty)-Q_\epsilon(\infty))+1-P_\epsilon(\infty)Q_\epsilon(\infty)\right]\text{.}
\end{multline*}
If there exist $\bar{T}_v(s_1,s_2)$, $\bar{T}_c(s_1,s_2)$ and $\bar{T}_r(s)$ such that
\begin{align*}
|T_v(t,s_1,s_2)| &\le \bar{T}_v(s_1,s_2),& \text{for all } t\\
|T_c(t,s_1,s_2)| &\le \bar{T}_c(s_1,s_2),& \text{for all } t\\
|T_r(t,s)| &\le \bar{T}_r(s),& \text{for all } t
\end{align*}
and such that
\begin{align*}
\lim_{t\to\infty}\sum_{s_1=1}^{t-1}\sum_{s_2=2s_1+1}^{2t} \bar{T}_{v}(s_1,s_2) < \infty\\
\lim_{t\to\infty}\sum_{s_1=0}^{t-1}\sum_{s_2=2s_1+2}^{2t} \bar{T}_{c}(s_1,s_2) < \infty\\
\lim_{t\to\infty}\sum_{s=1}^{2t}\bar{T}_{r}(s) < \infty
\end{align*}
then Theorem \ref{limbp} is a consequence of Lebesgue's dominated convergence theorem.
If $\epsilon\lambda'(P_\epsilon(\infty))\rho'(1-Q_\epsilon(\infty))<1$,
there exists $\delta>0$ such that 
\begin{equation*}
\epsilon(\lambda'(P_\epsilon(\infty))+\delta)(\rho'(1-Q_\epsilon(\infty))+\delta)<1.
\end{equation*}
On the other hand,
\begin{align}
|\lambda'(P_\epsilon(t))-\lambda'(P_\epsilon(\infty))|<\delta\label{cond1}\\
|\rho'(1-Q_\epsilon(t))-\rho'(1-Q_\epsilon(\infty))|<\delta\label{cond2}
\end{align}
for all but finite $t$.
One can therefore take $\bar{T}_v(s_1,s_2)$, $\bar{T}_c(s_1,s_2)$ and $\bar{T}_r(s)$ satisfying the above conditions 
by replacing
$\lambda'(P_\epsilon(t))$ and $\rho'(1-Q_\epsilon(t))$ in $T_v(t,s_1,s_2)$, $T_c(t,s_1,s_2)$ and $T_r(t,s)$
with $\lambda'(P_\epsilon(\infty)) + \delta$ and $\rho'(1-Q_\epsilon(\infty))+\delta$, respectively,
and multiplying them with an appropriate constant in order to take into account the fact 
that the bounded number of $\lambda'(P_\epsilon(t))$ and $\rho'(1-Q_\epsilon(t))$ in $T_v(t,s_1,s_2)$, $T_c(t,s_1,s_2)$ and $T_r(t,s)$ 
do not satisfy \eqref{cond1} and \eqref{cond2}.
\end{IEEEproof}

\bibliographystyle{IEEEtran}
\bibliography{IEEEabrv,ldpc}

\end{document}